\documentclass[twocolumn]{aastex62}

\usepackage{graphicx}
\usepackage{amsmath}  
\usepackage{xcolor}    

\newcommand{\angstrom}{\textup{\AA}}
\newcommand\Msun{$M_{\odot}\;$}
 
\submitjournal{ApJ}

\shorttitle{Mass Metallicity Relationship of SDSS Star Forming Galaxies}
\shortauthors{Sextl et al.}

\begin{document}

\title{Mass Metallicity Relationship of SDSS Star Forming Galaxies: Population Synthesis Analysis and Effects of Star Burst Length, Extinction Law, Initial Mass Function and Star Formation Rate
}

\correspondingauthor{Eva Sextl}
\email{sextl@usm.lmu.de}

\author{Eva Sextl}
\affiliation{Universit\"ats-Sternwarte, Fakult\"at f\"ur Physik, Ludwig-Maximilians Universit\"at M\"unchen, Scheinerstr. 1, 81679 M\"unchen, Germany}
\author{Rolf-Peter Kudritzki}
\affiliation{Universit\"ats-Sternwarte, Fakult\"at f\"ur Physik, Ludwig-Maximilians Universit\"at M\"unchen, Scheinerstr. 1, 81679 M\"unchen, Germany}
\affiliation{Institute for Astronomy, University of Hawaii at Manoa, 2680 Woodlawn Drive, Honolulu, HI 96822, USA}
\author{H. Jabran Zahid}
\affiliation{Microsoft Research, 14820 NE 36th St, Redmond, WA 98052, USA}
\author{I-Ting Ho}
\affiliation{Max-Planck-Institute for Astronomy, K\"onigstuhl 17, D-69117 Heidelberg, Germany}
\

\begin{abstract}
  We investigate the mass-metallicity relationship of star forming galaxies by analysing the absorption line spectra of $\sim$200,000 galaxies in the Sloan Digital Sky Survey. The galaxy spectra are stacked in bins of stellar mass and a population synthesis technique is applied yielding metallicities, ages and star formation history of the young and old stellar population together with interstellar reddening and extinction. We adopt different lengths of the initial starbursts and different initial mass functions for the calculation of model spectra of the single stellar populations contributing to the total integrated spectrum.  We also allow for deviations of the ratio of extinction to reddening R$_V$ from 3.1 and determine the value from the spectral fit. We find that burst length and R$_V$ have a significant influence on the determination of metallicities whereas the effect of the initial mass function is small. R$_V$ values are larger than 3.1. The metallicities of the young stellar population agree with extragalactic spectroscopic studies of individual massive supergiant stars and are significantly higher than those of the older stellar population. This confirms galaxy evolution models where metallicity depends on the ratio of gas to stellar mass and where this ratio decreases with time. Star formation history is found to depend on galaxy stellar mass. Massive galaxies are dominated by stars formed at early times.
  
\end{abstract}


\section{Introduction}

Star forming galaxies show a tight relationship between total stellar mass and average metallicity, the mass-metallicity relationship (``MZR'', see for instance \citealt{Lequeux1979, Tremonti2004, Kudritzki2016, Bresolin2022}). This relationship while evolving and becoming steeper holds out to redshifts z larger than 3.3 (\citealt{ValeAsari2009, Zahid2014, Genzel2015, Sanders2020}). In view of the complicated interplay between complex processes such as gas accretion, star formation, nucleosynthesis and chemical evolution, stellar and galactic winds, dynamical evolution and merging the existence of a simple relationship such as the MZR seems to be a Rosetta stone to understand galaxy formation and evolution.  As a consequence, a large variety of approaches has been developed  to use the MZR as a constraint of galaxy formation and evolution, \citet{Dave2011b}, \citet{Yates2012}, \citet{Dayal2013}, \citet{Schaye2015}, \citet{Peng2015}, \citet{Spitoni2017b, Spitoni2017a}, \citet{Pantoni2019}, \citet{Lapi2020}, \citet{Spitoni2020}, \citet{Trussler2020}, \citet{Kudritzki2021a, Kudritzki2021b} to name only a few. However, there is an important problem. Many of the observational MZR results obtained are based on measurements of the strongest ionized interstellar medium (ISM) emission lines, which are then converted into oxygen abundances. These ``strong line methods'' depend heavily on the calibration method used. \citet{Kewley2008} in an emission line analysis of 50,000 star forming Sloan Digital Sky Survey (SDSS) galaxies demonstrated convincingly that the systematic uncertainties in oxygen abundances can be as large as 0.8 dex. As a result, the slope of the MZR can vary significantly for the same set of observed emission line data depending on the calibration applied.

An obvious alternative approach is the analysis of low resolution absorption line spectra of individual stars. Blue and red  supergiants (BSG and RSG) as the brightest stars at visual or infra-red light, respectively, are the ideal objects for this purpose. Galaxy metallicities obtained by this method have only small uncertainties of at most 0.1 dex (systematic and random, see \citealt {Gazak2015}) and have been used to construct a very reliable MZR, see \citealt{Kudritzki2016, Davies2017, Bresolin2022}, and references therein. Unfortunately, the method is limited with respect to galactic  distance. The most distant objects studied so far are the Antennae galaxies and their super star clusters at 20 Mpc \citep{Lardo2015}.

In view of this limitation a third technique which has been well advanced over the last two decades has become truly important, the analysis of stellar absorption line spectra of the integrated stellar population through population synthesis techniques. The advantage of this technique is that it reaches out to much larger distances because of the relatively high surface brightnesses of galaxies.  In addition,  it can provide information about the young and old stellar population and the star formation history.

The general idea was pioneered by \cite{Tinsley1968} and \cite{Spinrad1972}. With significant further improvements of stellar evolution theory and stellar spectral modelling the concept was then advanced by many groups, see for instance \cite{Arimoto1987}, \cite{Bressan1996}, \cite{Reichardt2001},  \cite{Bruzual2003}, \cite{Cardiel2003} and references therein. More recent work used the advantages of increased computer power and improved the method for applications on massive spectroscopic galaxy surves, first focussing on spectra of the central regions of galaxies (\citealt{Gallazzi2005, Cid_Fernandes2005, Panter2008, Conroy2009, Koleva2009, Sanchez2011,  Peng2015, Zahid2017}) and later on spectra of spatially resolved regions of galaxies obtained with Integral Field Units (\citealt{Yoachim2012, Gonzalez2015, Schaefer2017, Parikh2021}).

Given the enormous potential of this technique it is important to carefully assess the systematic effects arising from the large variety of assumptions which enter this approach.  While this has been done already to a large extend in previous work, our work presented here is investigating a variety of additional important aspects, which have not been addressed so far in much detail. We study the effects of finite star formation burst lengths compared with instantaneous bursts, the effects of pre-main sequence evolution and initial mass function and we assess the influence of an interstellar reddening law characterized by a ratio R$_V$ of extinction to reddening deviating from the standard value of 3.1. We will also investigate the influence of star formation rate (SFR) on MZR, extinction and ages. For this purpose we start from our previous work \citep{Zahid2017} and apply an improved technique to analyse a sample of 200,000 SDSS star forming galaxies with spectra stacked into bins of stellar mass. We compare the MZRs obtained in this way for the different assumptions with the MZR from supergiant spectroscopy, which we use as benchmark. We will also discuss stellar ages and star formation history.

\begin{figure}[ht!]
  \begin{center}
    \includegraphics[scale=0.50]{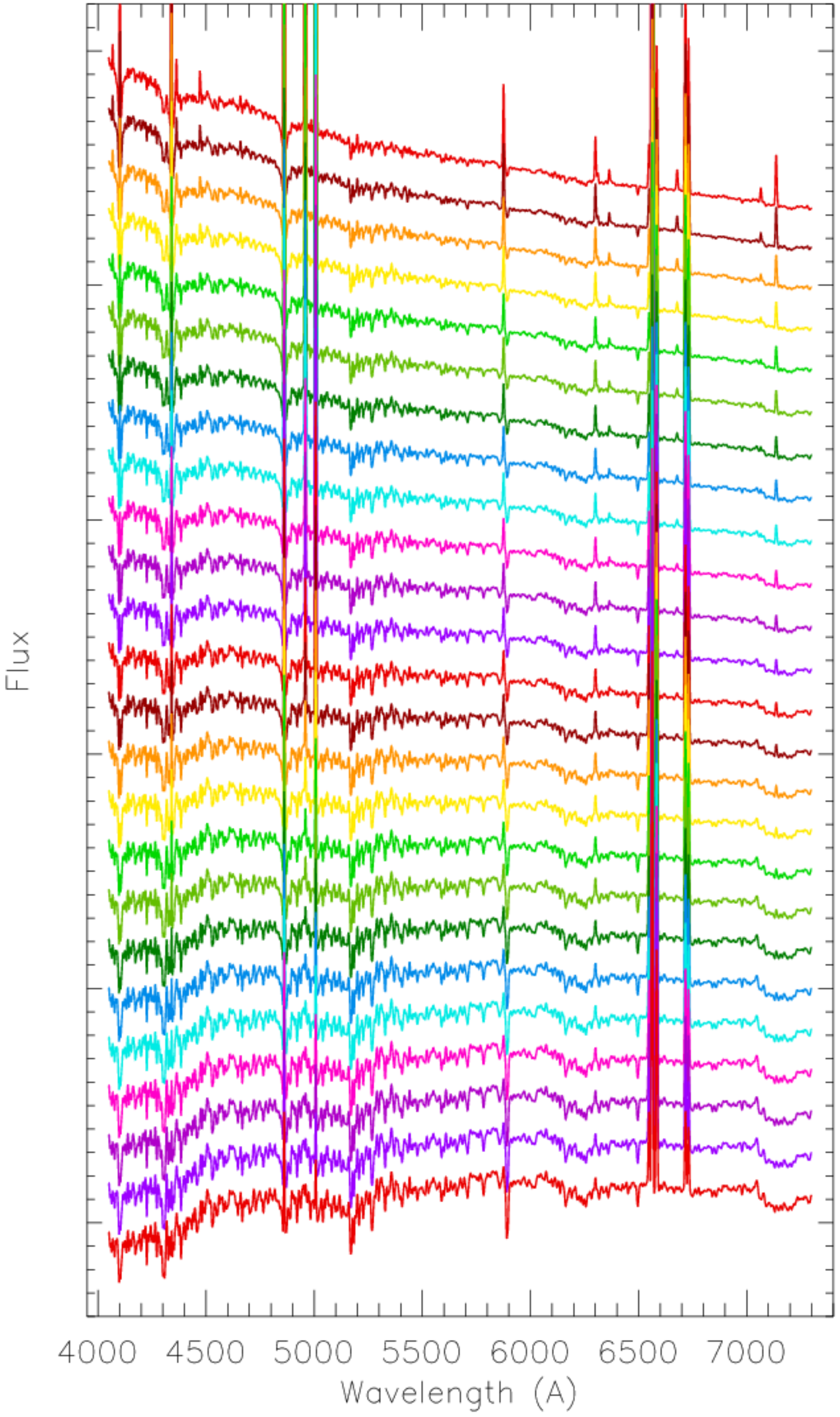}
  \end{center}
  \caption{
Fluxes (arbitrary units) of the stacked SDSS spectra in 25 mass bins from log M$_*$ = 8.55 (top) to 10.95 (bottom). For clarity the spectra are shifted by 0.2 units from one mass bin to the next.
} \label{figure_1}
\end{figure}

 \begin{figure}[ht!]
  \begin{center}
    \includegraphics[scale=0.47]{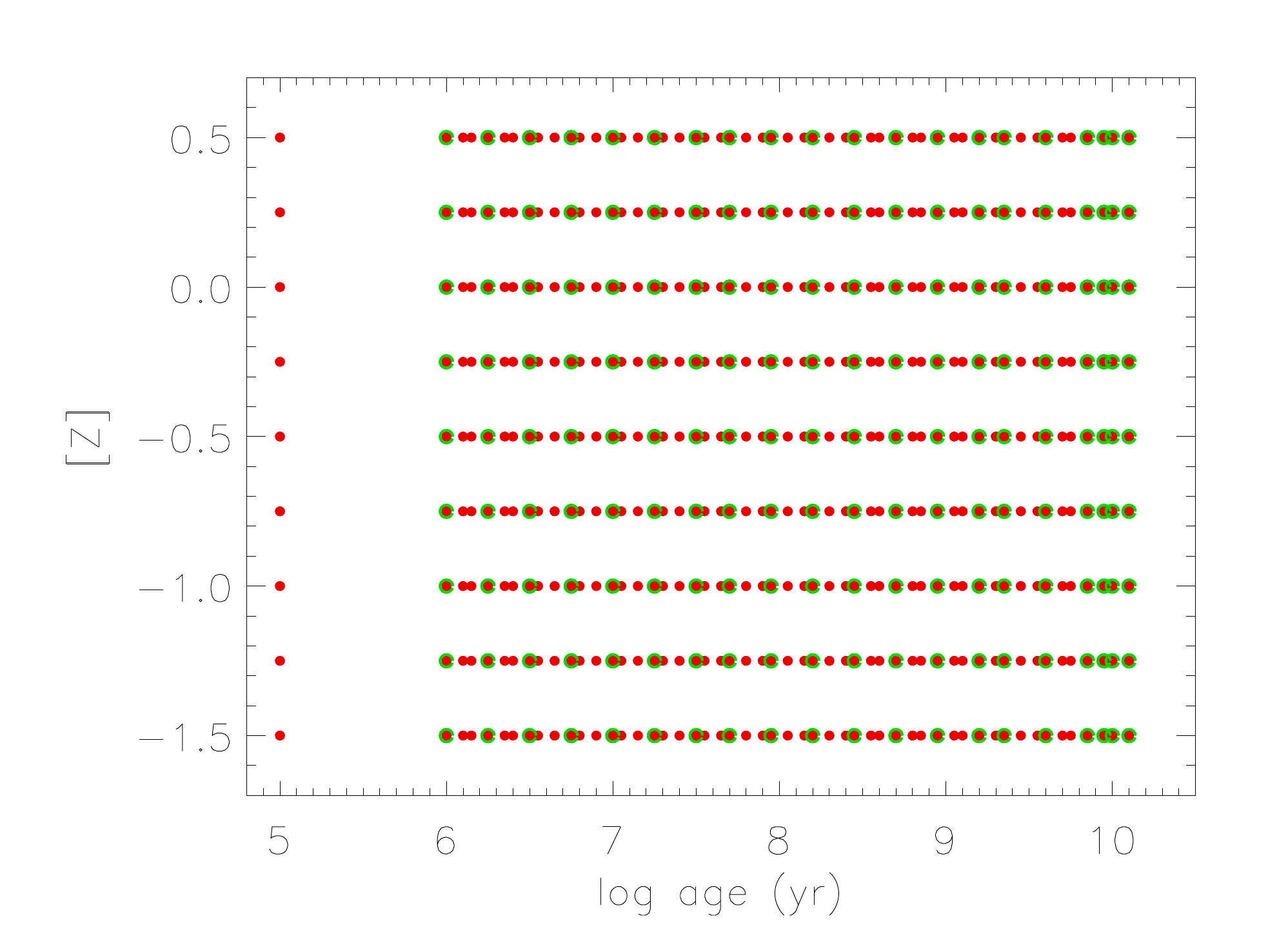}
  \end{center}
  \caption{
The grid of SSB metallicities [Z] and ages for the calculation of model spectra using eq. (1). Filled red circles: high age resolution grid, green circles: low resolution grid.
} \label{figure_2}
\end{figure}

\section{Observed Spectra and Analysis Technique}

The same dataset as described in \cite{Zahid2017} is used. The observations consist of SDSS spectra of about 200,000 star forming galaxies in the redshift range 0.027 $\le$ z $\le$ 0.25. The galaxies were selected according to the strengths of their interstellar medium (ISM) emission lines and the redshift range was chosen to ensure rest-frame wavelength coverage of the [OII] $\lambda$3727 and $\lambda$7330 emission lines. Since we are interested only in the average galaxy properties as a function of their stellar mass, we have stacked the spectra in 25 bins of stellar mass log M$_*$ (measured in units of \Msun \hspace{-3pt}) from 8.55 to 10.95 in steps of 0.1 dex. In this way the signal-to-noise is strongly enhanced for the analysis and ranges from 160 at the low mass end to 1300 at higher masses (see Table in \cite{Zahid2017}). To avoid systematic effects as a function of stellar mass caused by different line widths we have convolved all spectra to the line width FWHM = 330 km s$^{-1}$ found for the bin with the highest stellar mass. In addition, the spectra are normalized by setting the mean flux between 4400 and 4450 \angstrom\ equal to unity. Details of galaxy selection, stellar mass determination, stacking and convolution procedure are given in \cite{Zahid2017}. 

The spectra of the 25 mass bins are shown in  Figure \ref{figure_1}. Strikingly, the SED slope changes from low to high stellar mass and the stellar absorption lines become stronger. We will show that this is caused by systematic differences in interstellar extinction, average stellar age and metallicity.

The technique to analyse these spectra in order to extract information about stellar population metallicity and ages is a modification of the approach introduced by \cite{Zahid2017}. We model the spectrum of the integrated stellar population M$_{\lambda}$ as a linear combination of the spectra of single stellar bursts (SSB) f$_{\lambda, i}$(t$_{i}$, [Z]$_{i}$)  with age t$_{i}$ and logarithmic metallicity [Z]$_{i}$ = log Z/Z$_{\sun}$

\begin{equation}
  M_{\lambda} = D_{\lambda}(R_{V}, E(B-V)) \sum_{i=1}^{n_{SSB}} b_{i} f_{\lambda, i} (t_{i}, [Z]_{i}).
\end{equation}  

The coefficients b$_i$ describe the contribution of burst i with age t$_{i}$ and metallicity  [Z]$_{i}$.
D$_{\lambda}(R_{V}, E(B-V))$ accounts for the absorption by interstellar dust, which depends on interstellar reddening E(B-V) and R$_V$ = A$_V$/E(B-V) the ratio of interstellar extinction to reddening in the V-band. The value of R$_V$ characterizes the reddening law, at least to first order. As is well known from studies of stellar SEDs in star forming regions,  R$_V$ can cover a wide range from 2 to 6 (see, for instance, \citealt{urbaneja2017}). We use the interstellar reddening law by \cite{Odonnell1994}, which is a modification of \cite{Cardelli1989}.


\begin{figure}[ht!]
  \begin{center}
    \includegraphics[scale=0.45]{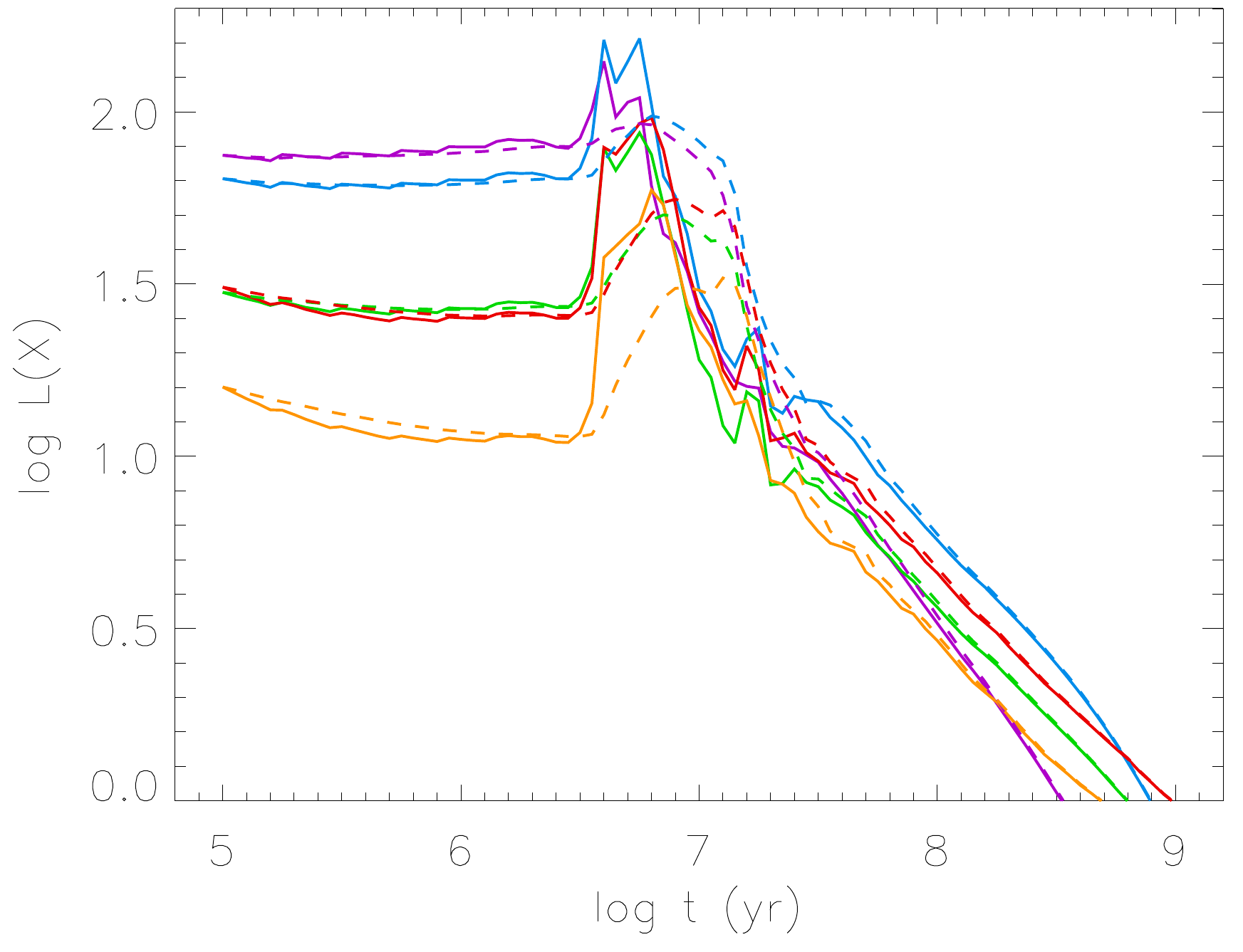}
    \includegraphics[scale=0.45]{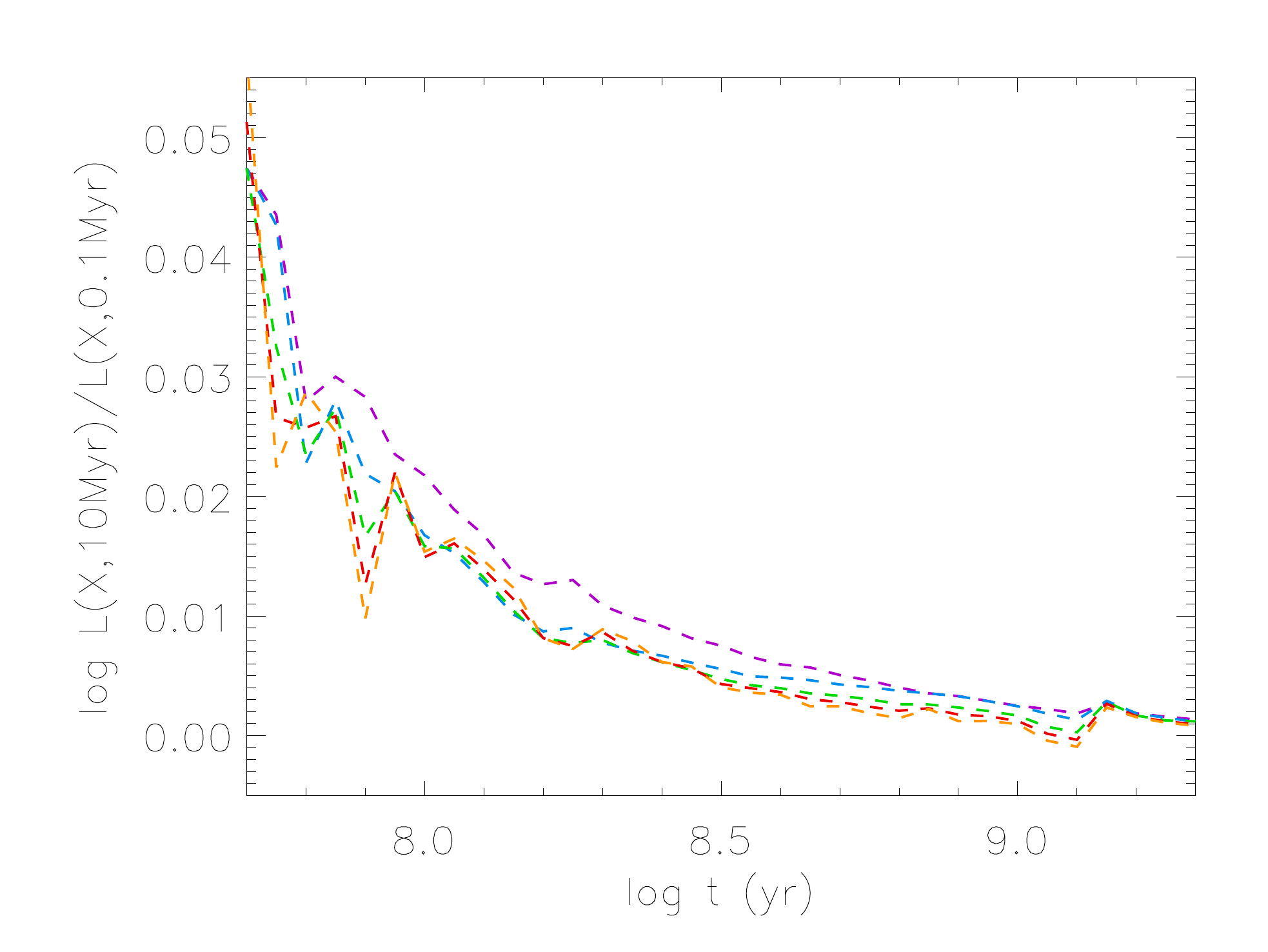}
  \end{center}
  \caption{
    Luminosities of SSBs as a function of age (yrs) for different spectral passbands (pink: U, blue: B, green: V, red: R, orange: I). The solid curves correspond to a burst duration t$_b$ = 0.1 Myr, the dashed curves to t$_b$ = 10 Myr. The bottom figure shows the logarithm of the ratio of luminosities of 10 Myr to 0.1 Myr bursts at larger ages. A metallicity of [Z]=0 has been used for this example.   
} \label{figure_3}
\end{figure}

\begin{figure}[ht!]
  \begin{center}
    \includegraphics[scale=0.45]{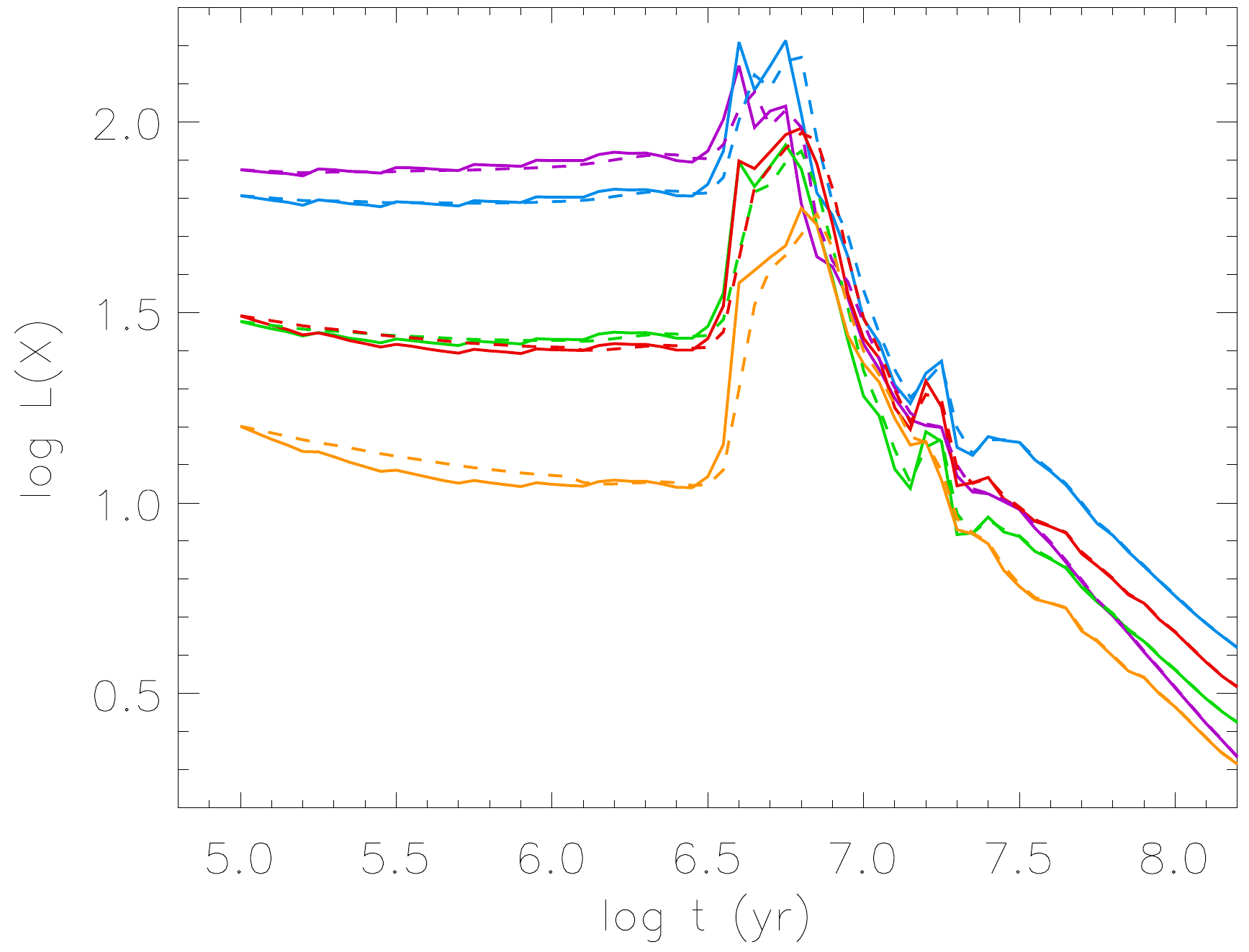}
  \end{center}
  \caption{
Same as Figure \ref{figure_3} (top) but the dashed curves correspond to t$_b$ = 1 Myr.
} \label{figure_4}
\end{figure}

The SSB spectra are calculated using the Flexible Stellar Population Synthesis code (FSPS, version 3.2) \citep{Conroy2009, Conroy2010} and adopting a \cite{Chabrier2003} initial mass function, the MILES stellar spectral library  \citep{Sanchez2006} and the MESA stellar evolution isochrones (MIST, \citealt{Choi2016, Dotter2016}). We also account for finite time lengths of the stellar bursts (0.1, 1.0, 10.0 Myr). This will be discussed in the next section. Finally, our model spectra are normalized in the same way as the observed spectra, convolved to the equal resolution and interpolated to the same integer values of wavelength points as the observed spectra. 

The modelling of the spectra by eq. (1) does not account for the contribution of a potential HII-region nebular continuum, which can become important in cases of extremely high star formation rates with very strong nebular emission as shown by \cite{Cardoso2022}. However, the strength of the hydrogen nebular emission lines relative to the stellar continuum allows us to estimate the effects of nebular continuum emission for our analysis. The H$_{\beta}$ emission equivalent widths in our stacked spectra is $\lesssim$ 20 \AA~and, thus, the nebular continuum contribution to the total spectrum is very small. We have verified this conclusion by estimating the nebular contribution from the measurement of H$_{\beta}$ emission equivalent width and by subtracting a theoretical reddened nebular continuum. We have then repeated the analysis with modified spectrum and found only very small changes with respect to metallicity ($\leq$ 0.01 dex) and age ($\leq$ 1\%).  In our analysis of the SFR dependence of the stellar MZRs in section 9 below we have encountered a very few cases in the highest SFR quintile with larger H$_{\beta}$ equivalent widths. But even here the effects on metallicity were smaller than 0.05 dex.

\begin{figure}[ht!]
  \begin{center}
    \includegraphics[scale=0.45]{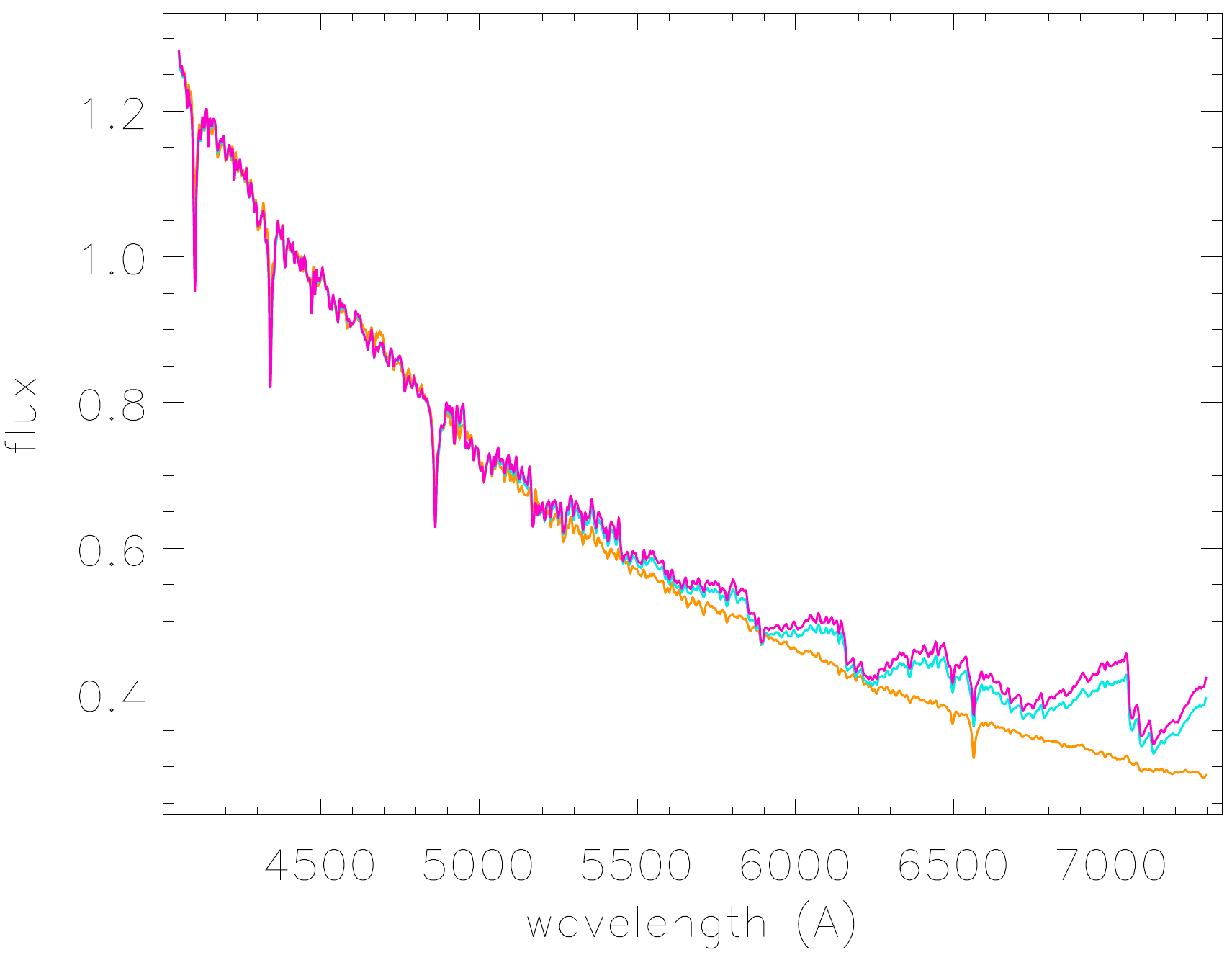}
    \includegraphics[scale=0.45]{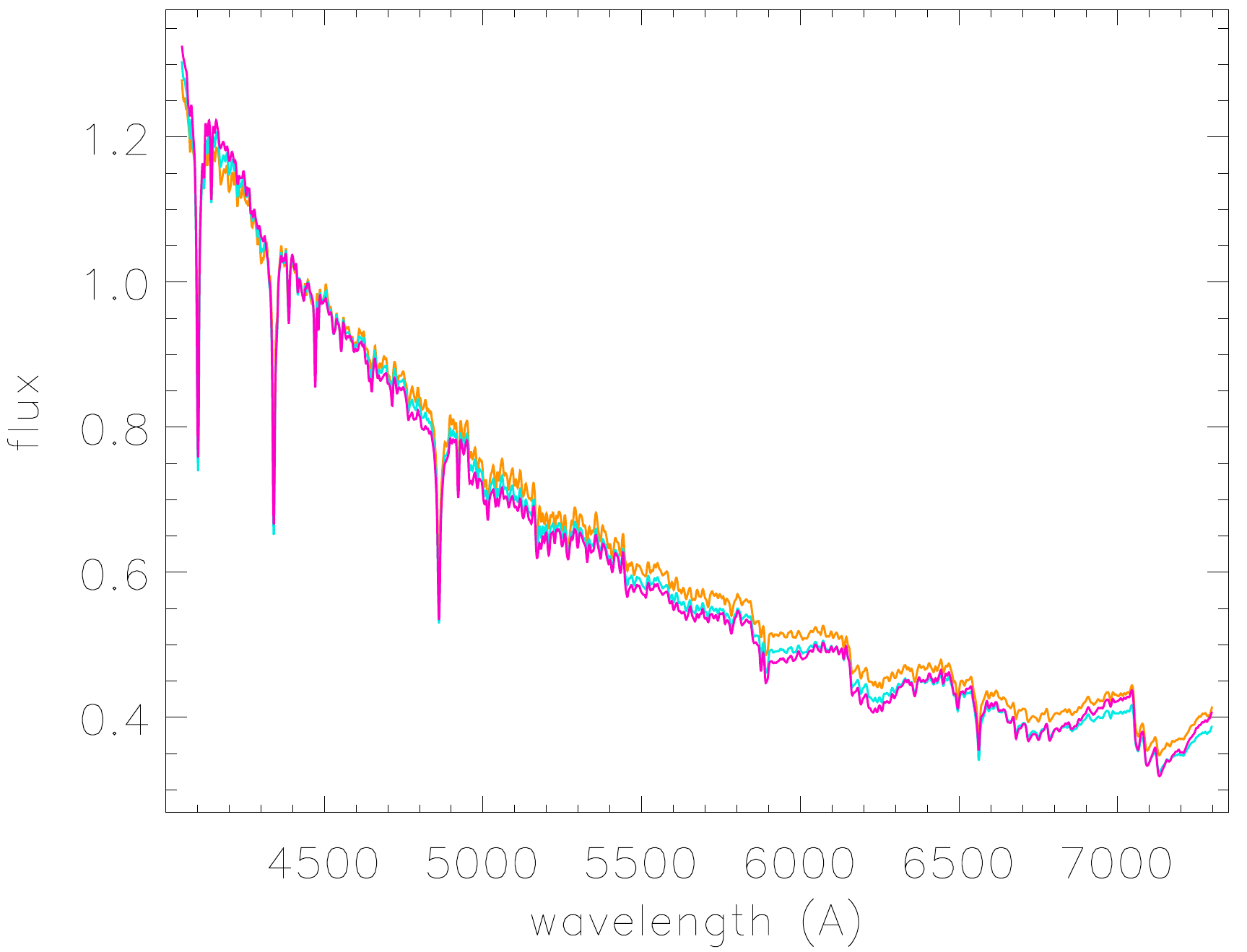}
    \includegraphics[scale=0.45]{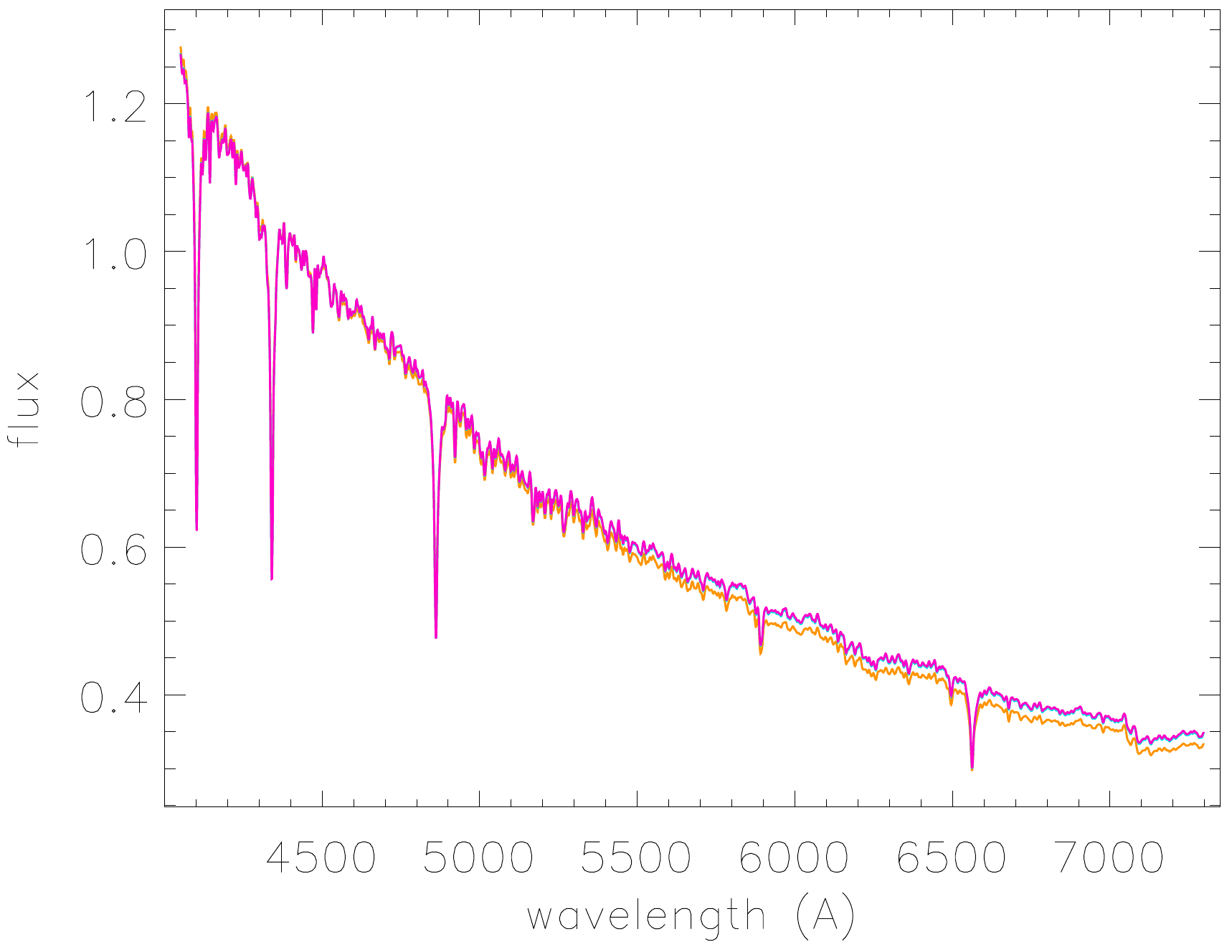}
  \end{center}
  \caption{
    Spectra of SSBs at different ages for different burst lengths (pink:  t$_b$ = 0.1 Myr, cyan: 1 Myr, orange: 10 Myr). Top: age = 10 Myr, middle: 19 Myrs, bottom: 45 Myrs)   
} \label{figure_5}
\end{figure}

For the selection of metallicities and ages of the SSB spectra we use two rectangular grids consisting of 9 [Z] values and 52 (high age resolution)  and 20 (low age resolution) ages, respectively. The metallicities range from [Z] = -1.5 to 0.5 increasing in steps of 0.25 dex. The low resolution ages start at 1Myr and increase in logarithmic steps of $\Delta$ log t[Gyr] = 0.25 until 7.08 Gyr and then continue with 8.91, 10.0 and 12.59 Gyr. The high resolution age grid starts at 0.1 Myr with a step to 1.0 Myr and then continues with logarithmic steps $\Delta$ log t[Gyr] alternating between 0.05 and 0.1 until 12.59 Gyr are reached.  This leads to a total number of SSB n$_{SSB}$ = 468 in the high age resolution and 180 in the low resolution case. Metallicities and ages are illustrated in Figure \ref{figure_2}.

In the fit of the observed spectra with the model spectra we then adopt a grid of R$_V$ and E(B-V) values. For each pair of R$_V$ and E(B-V) we calculate the wavelength dependent dust attenuation factor  D$_{\lambda}(R_{V}, E(B-V))$ and use the bonded-variables least squares (BVLS) algorithm by \cite{Stark1995}, translated to IDL by Michele Cappellari\footnote{\url{https://www-astro.physics.ox.ac.uk/~cappellari/idl/bvls.pro}} to determine the coefficients b$_i$. With the b$_i$ the comparison of the model spectrum with the observations leads to a $\chi^2$ value for each R$_V$, E(B-V) pair and the minimum of $\chi^2$ defines the best fit yielding final values for R$_V$ and E(B-V) and average metallicity [Z] and age t$_{av}$ defined as

\begin{equation}
  [Z] = \sum_{i=1}^{n_{SSB}} b_{i} [Z]_{i}
\end{equation}

and

\begin{equation}
  t_{av} = \sum_{i=1}^{n_{SSB}} b_{i} t_{i}.
\end{equation}

Introducing a young and old stellar population through t$_i \le$ 1.6 Gyr and t$_i \ge$ 1.6 Gyr, respectively, we can calculate corresponding metallicities [Z]$_{young}$, [Z]$_{old}$ and ages t$_{young}$, t$_{old}$ via

\begin{equation}
  b_{young} = \sum_{i_{young}} b_{i} ,  b_{old} = \sum_{i_{old}} b_{i}
\end{equation}  

and

\begin{equation}
  [Z]_{young} = {1 \over b_{young}} \sum_{i_{young}} b_{i} [Z]_{i}
  \end{equation}
  \begin{equation}
  [Z]_{old} = {1 \over b_{old}} \sum_{i_{old}} b_{i} [Z]_{i}
  \end{equation}
  \begin{equation}
  t_{young} = {1 \over b_{young}} \sum_{i_{young}} b_{i} t_{i}
  \end{equation}
  \begin{equation}
  t_{old} = {1 \over b_{old}} \sum_{i_{old}} b_{i} t_{i}.
\end{equation}  

The choice of 1.6 Gyr for the definition of the young population is motivated by galaxy chemical evolution models which show only small changes ($\leq$ 0.1 dex) of metallicity during the final 1 to 2 Gyrs of evolution \citep{Spitoni2017b, Spitoni2017a, Kudritzki2021a, Kudritzki2021b}. Moreover, the molecular gas depletion time which defines star formation efficiency, is of the same order (1.9 Gyr, see \citealt{Leroy2008}). They young and old population fitting is insensitive to the selected value, for instance, chosing 2.5 instead of 1.6 Gyr does not change the results.

We note that due to the normalization of our spectra between 4400 and 4450 \AA~the [Z] and t-values obtained in this way are effectively B-band luminosity weighted metallicities and ages. This will become more evident further below.

The results for metallicities, ages, reddening and extinction obtained in this way will be discussed in the next sections.

\begin{figure}[ht!]
  \begin{center}
    \includegraphics[scale=0.50]{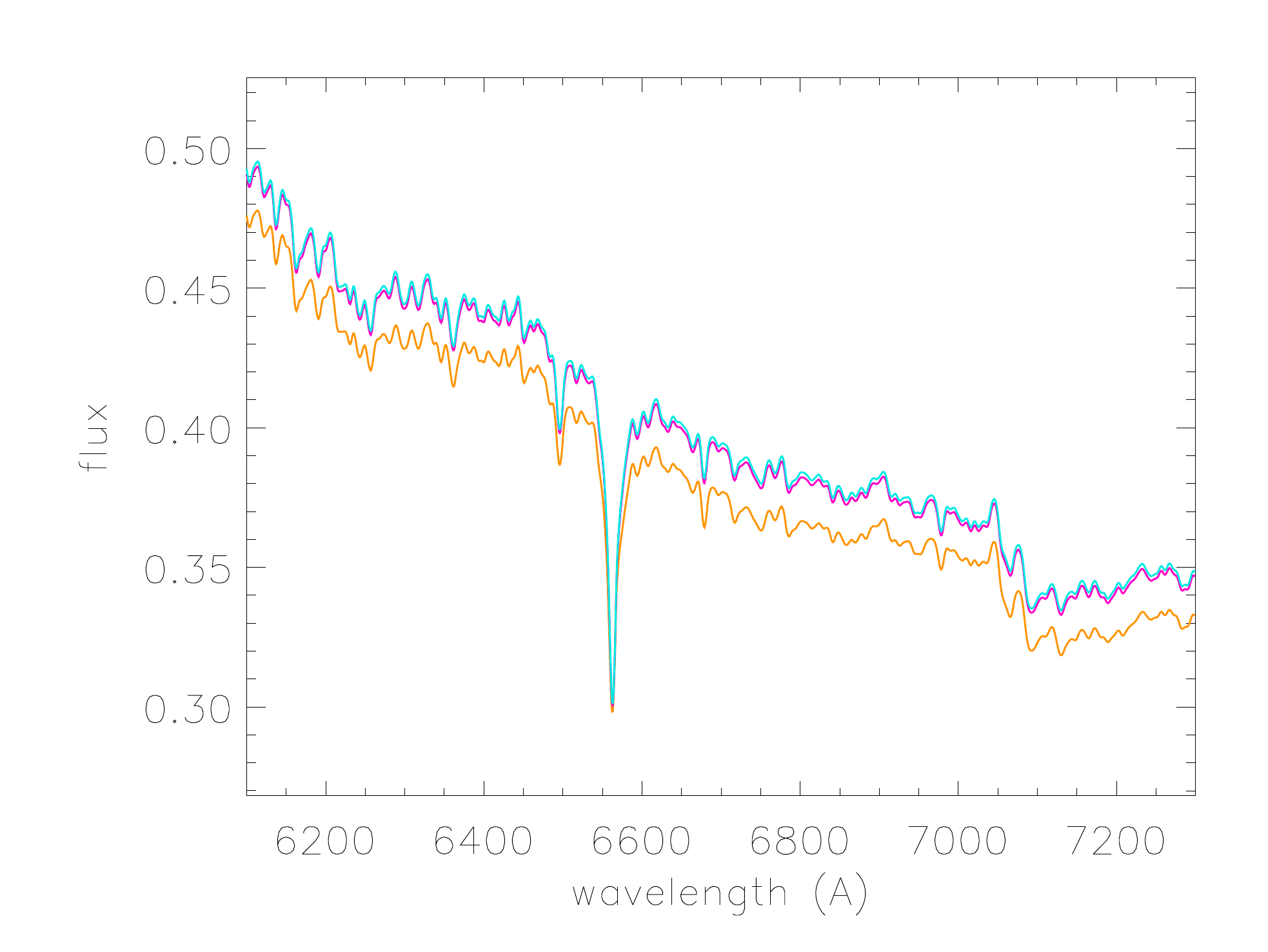}
    \includegraphics[scale=0.50]{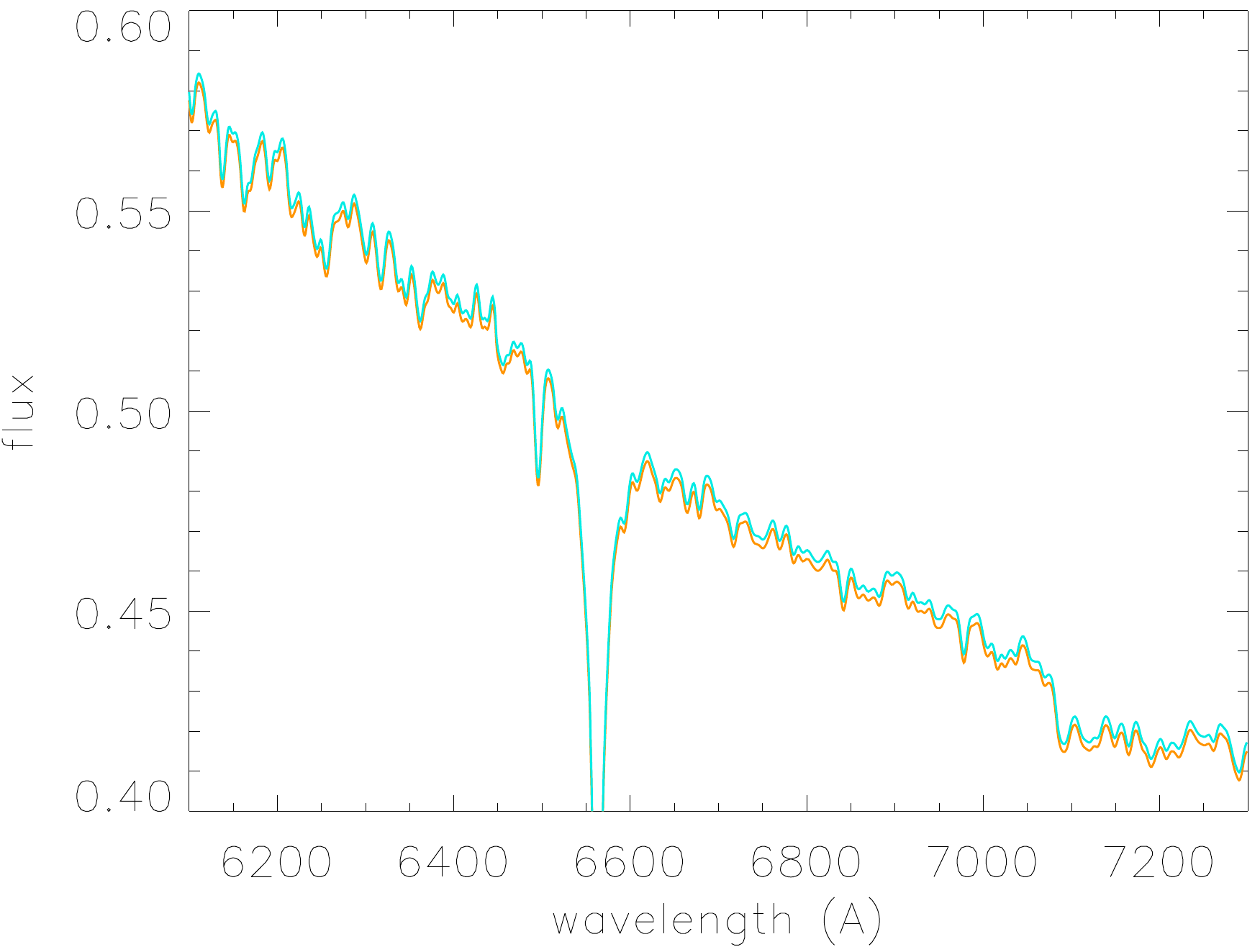}
  \end{center}
  \caption{
Same as Figure \ref{figure_5} but with a different wavelength range for ages of 45 Myr (top) and 501 Myr (bottom).
} \label{figure_6}
\end{figure}

\section{The Effects of Star Formation Burst Length on SSB spectra}

Most commonly, the spectra of star bursts are calculated by adopting the approximation of so-called ``instantaneous bursts'', for which the stellar evolution isochrone at one singular time step is used. However, as is well known, star formation in star forming regions is not instantaneous but happens over finite time scales. For instance, as shown by \cite{Efremov1998}, the star formation time t$_b$ of a burst is correlated with the size of the star forming region and can cover a wide range from 0.1 to 10 Myr in regions where associations and star clusters form. More recent work on star formation in molecular clouds confirmed burst lengths of a few million years \citep{Corbelli2017, Kim2021}. It is, thus, important to investigate what influence t$_b$ can have on the burst model spectra and the resulting analysis.

If L$_{iso}$($\lambda,\tau$) is the total luminosity per Angstroem at wavelength $\lambda$ of an isochrone of age $\tau$, then the luminosity L($\lambda$,t,t$_0$,t$_b$) of a burst at time t, which started at t$_0$  is

\begin{equation}
  L(\lambda,t,t_0,t_b) = {1 \over (t-t_0)} \int_{t_0}^{t} L_{iso}(\lambda,\tau)d\tau,
\end{equation}  
\\
for t$_{0} \le$ t $\le$ t$_{b}$+t$_{0}$ and

\begin{equation}
  L(\lambda,t,t_0,t_b) = {1 \over t_b} \int_{t-t_0}^{t} L_{iso}(\lambda,\tau)d\tau
\end{equation}  
\\
for t $\ge$  t$_{b}$+t$_{0}$.\\

We adopted constant star formation rate during the burst and L($\lambda$,t,t$_0$,t$_b$) is the luminosity averaged over the duration of the burst. Figures \ref{figure_3} and \ref{figure_4} show the effects of burst duration on the luminosities of SSBs as a function of age in different spectral bands. They are quite dramatic around the peak at ages of ~5-10 Myrs (up to 0.3 dex) and then become weaker but continue to be noticeable ($\geq$ 0.005 dex) compared with the S/N of the spectra until values of t$_b$/t $\approx$ 0.02 are reached (see Figures \ref{figure_3} bottom). This is a consequence of the power law behaviour L$_{iso}$(X,t)$ \propto$ t$^{-n}$ for larger values of t which leads to L(X,t,t$_0$,t$_b$) = L$_{iso}$(X,t)(1+nt$_b$/(2t)) for small values of t$_b$/t. X stands for the spectral passband (U, V, B, R, I). In the same way the spectra of SSBs show the effects of burst length. Examples are given in figures \ref{figure_5} and \ref{figure_6}. The differences between the spectra of different burst length become small at ages much larger than the burst length as discussed above. We note that all luminosities are given in units of the solar luminosity L$_\sun$.

The influence of the adopted burst lengths on the analysis of the observed spectra will be investigated in the next section.

\begin{figure}[ht!]
  \begin{center}
    \includegraphics[scale=0.45]{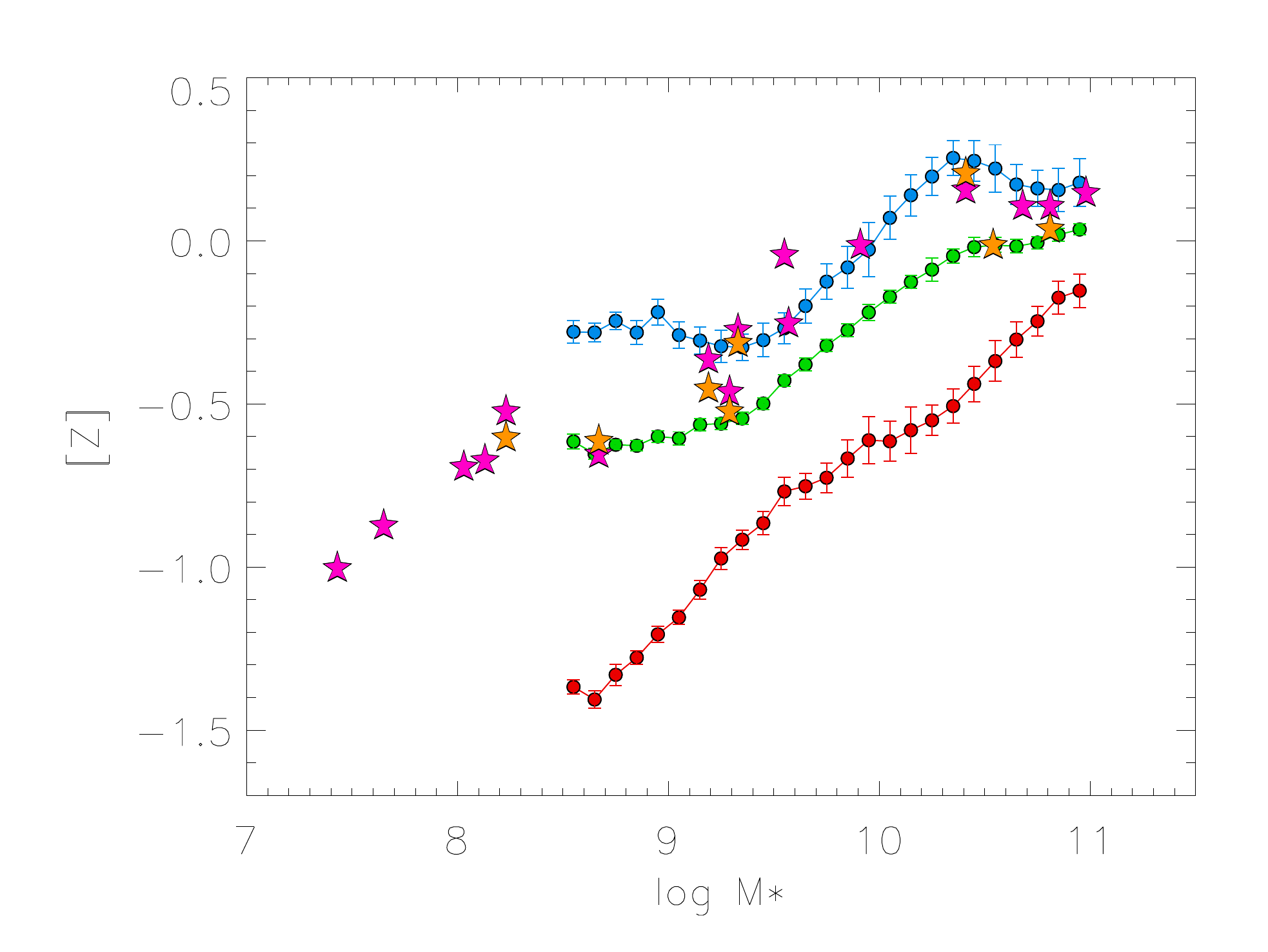}
    \includegraphics[scale=0.45]{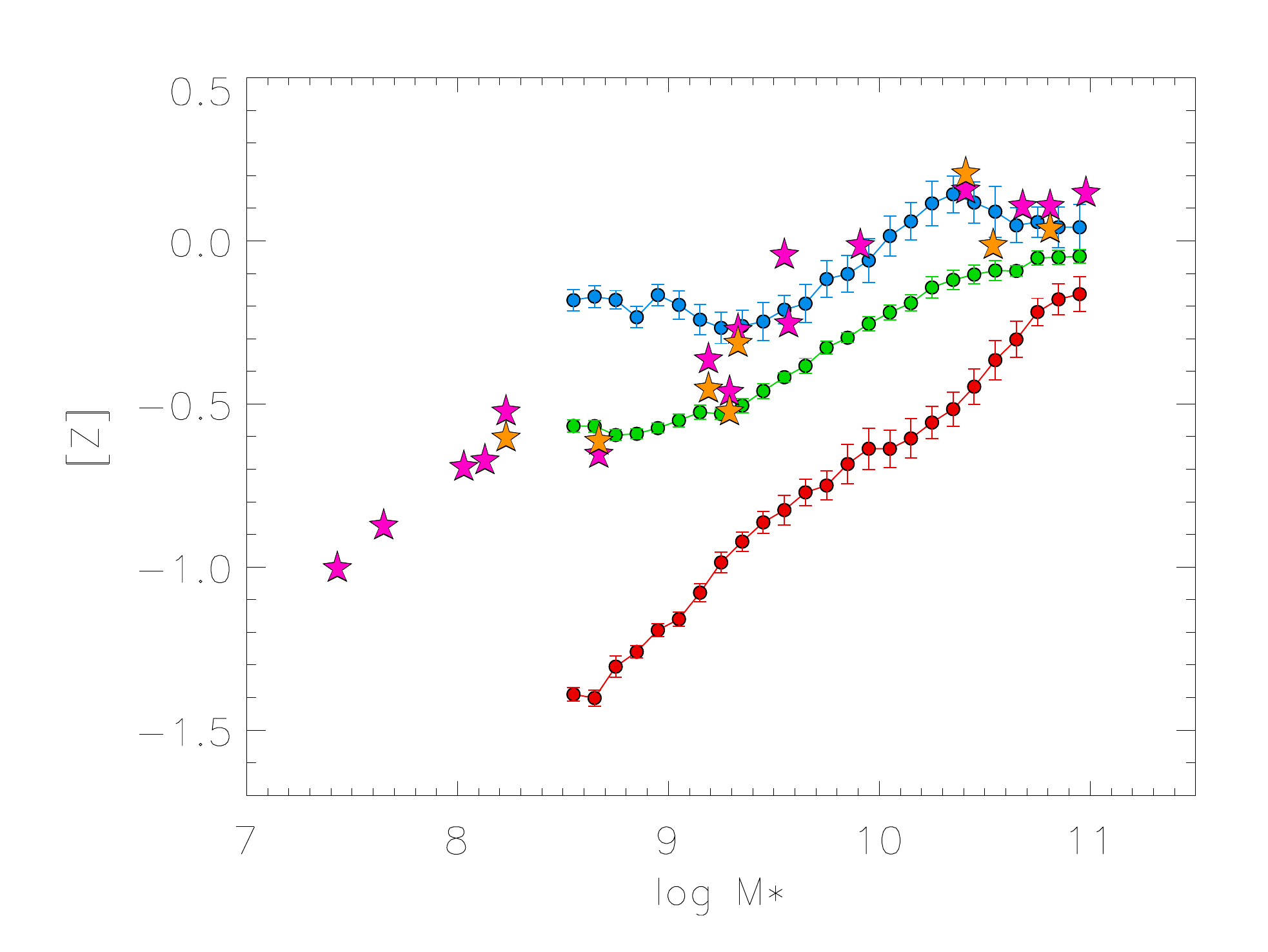}
    \includegraphics[scale=0.45]{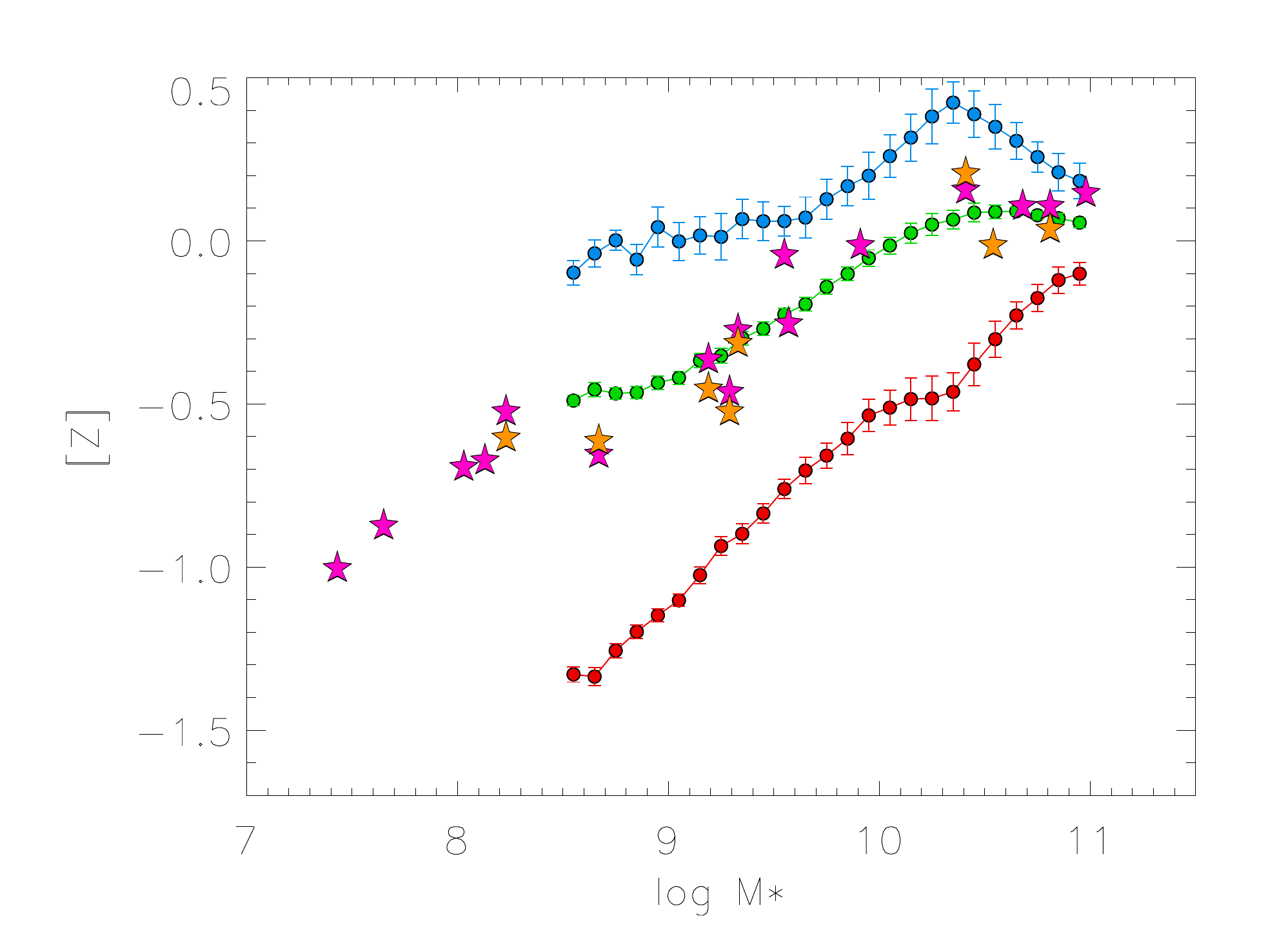}
  \end{center}
  \caption{Analysis of SDSS star forming galaxies. Metallicities of the young (blue) and old (red) stellar population and average metallicities (green). The grid of SSB with high time resolution was used. Top: 0.1 Myr burst length, middle: 1 Myr Myr, bottom: 10 Myr. The MZR of blue and red supergiant stars (pink and orange stars) obtained from the analysis of individual stars in 17 galaxies is also shown as a benchmark for the young stellar population.
} \label{figure_7}
\end{figure}

\begin{figure}[ht!]
  \begin{center}
    \includegraphics[scale=0.45]{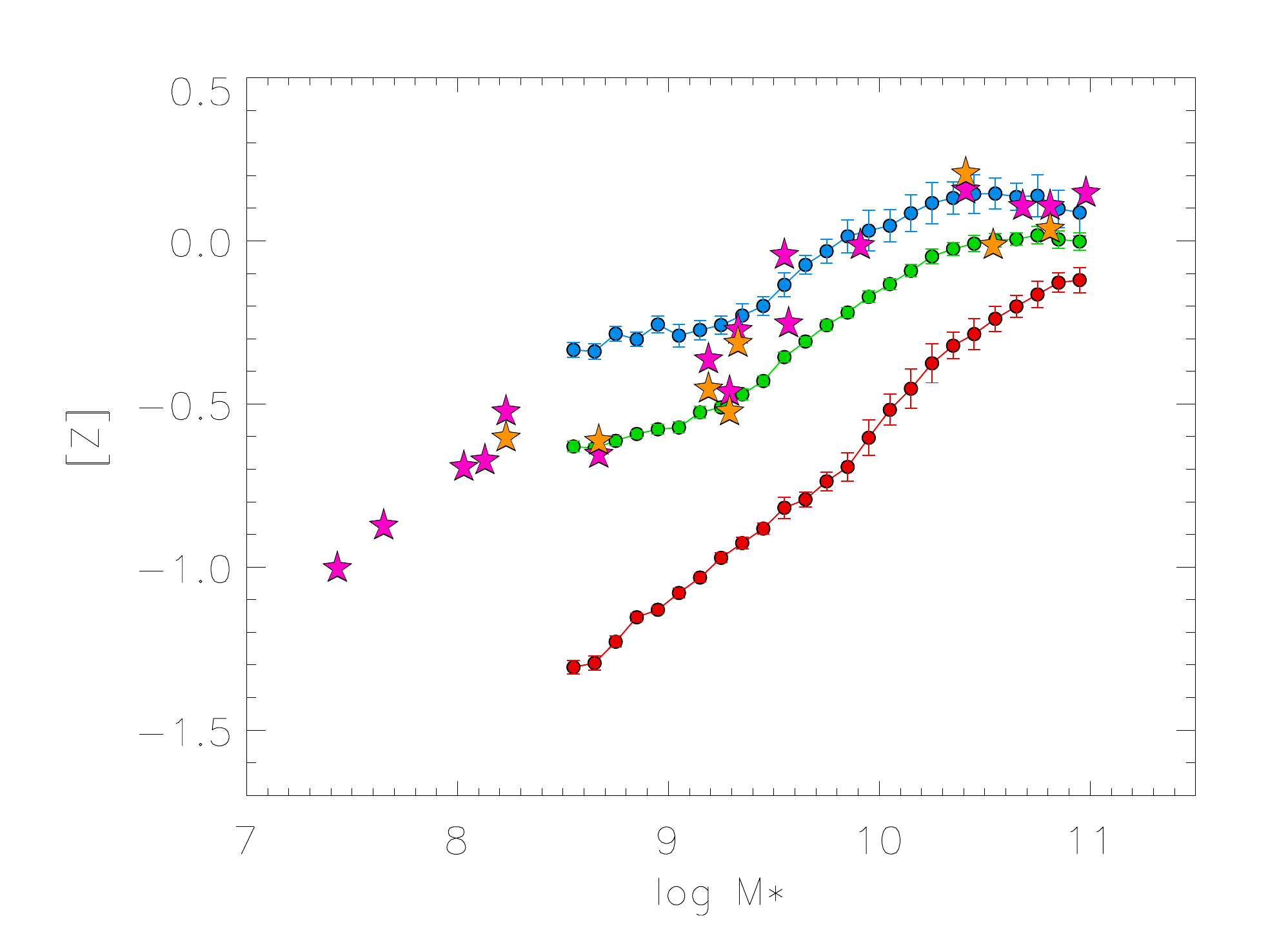}
    \includegraphics[scale=0.45]{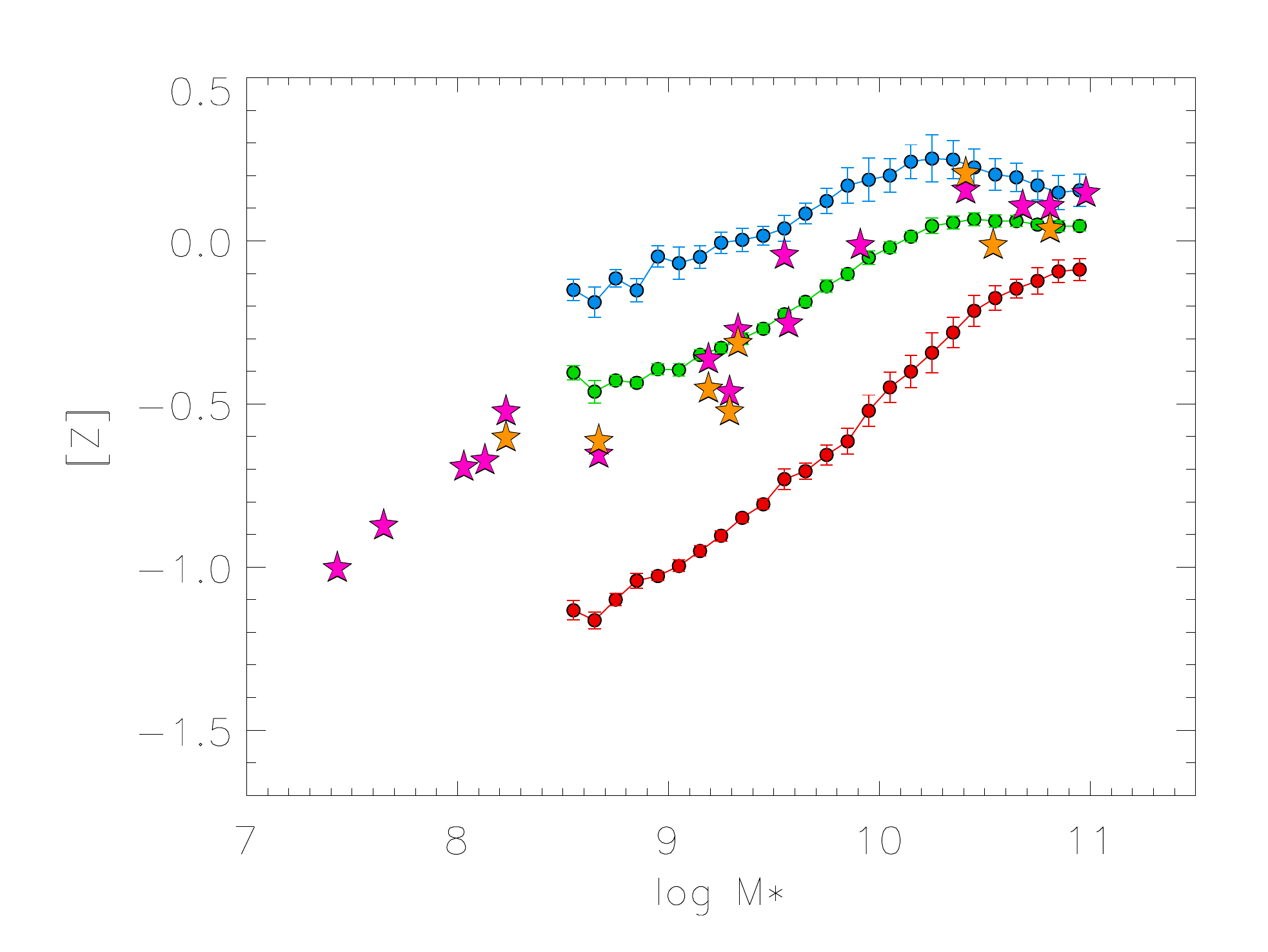}
    \includegraphics[scale=0.45]{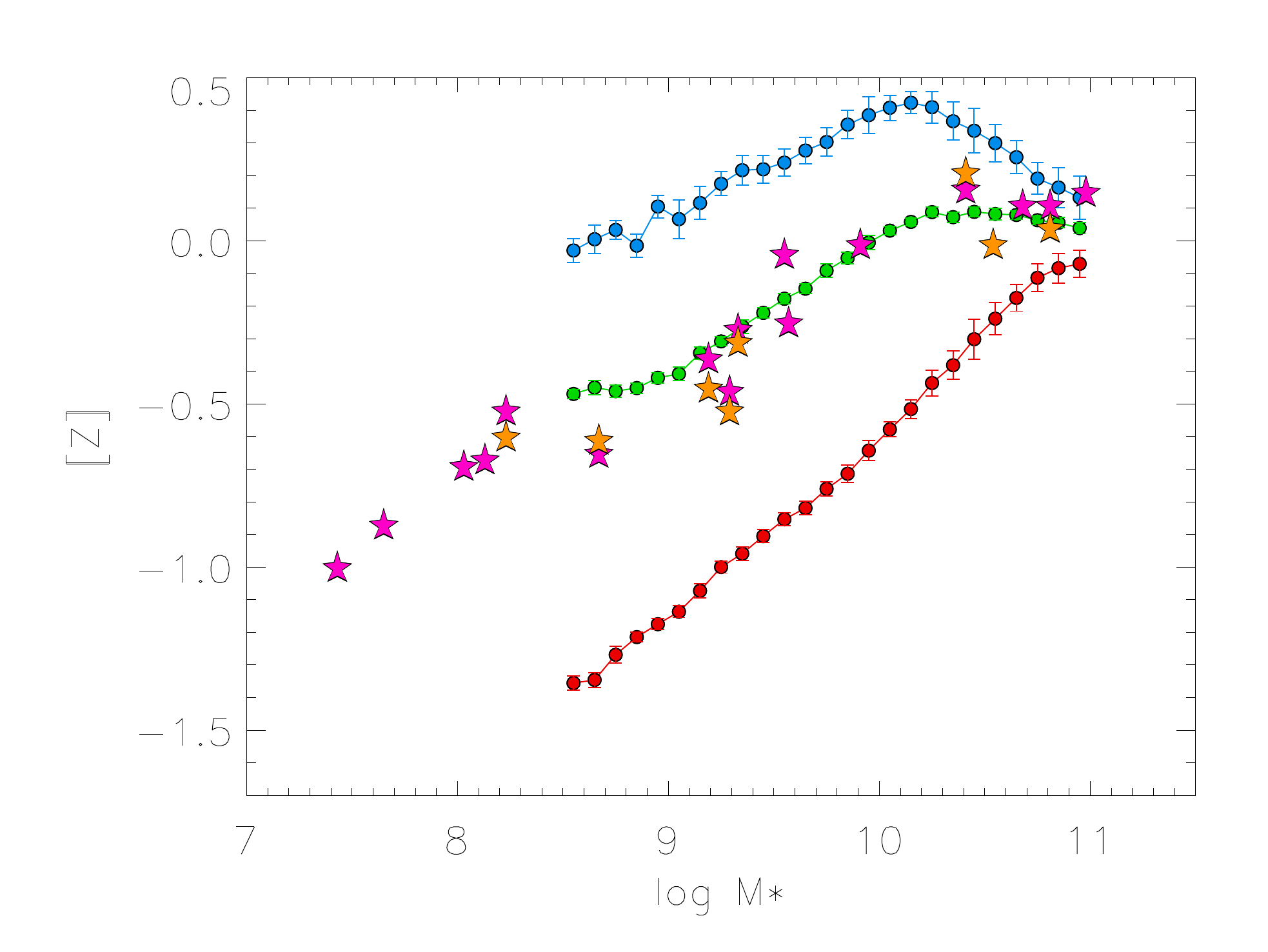}
  \end{center}
  \caption{
Same as Figure \ref{figure_7} but for the SSB grid with low time resolution.
} \label{figure_8}
\end{figure}

\begin{figure}[ht!]
  \begin{center}
    \includegraphics[scale=0.40]{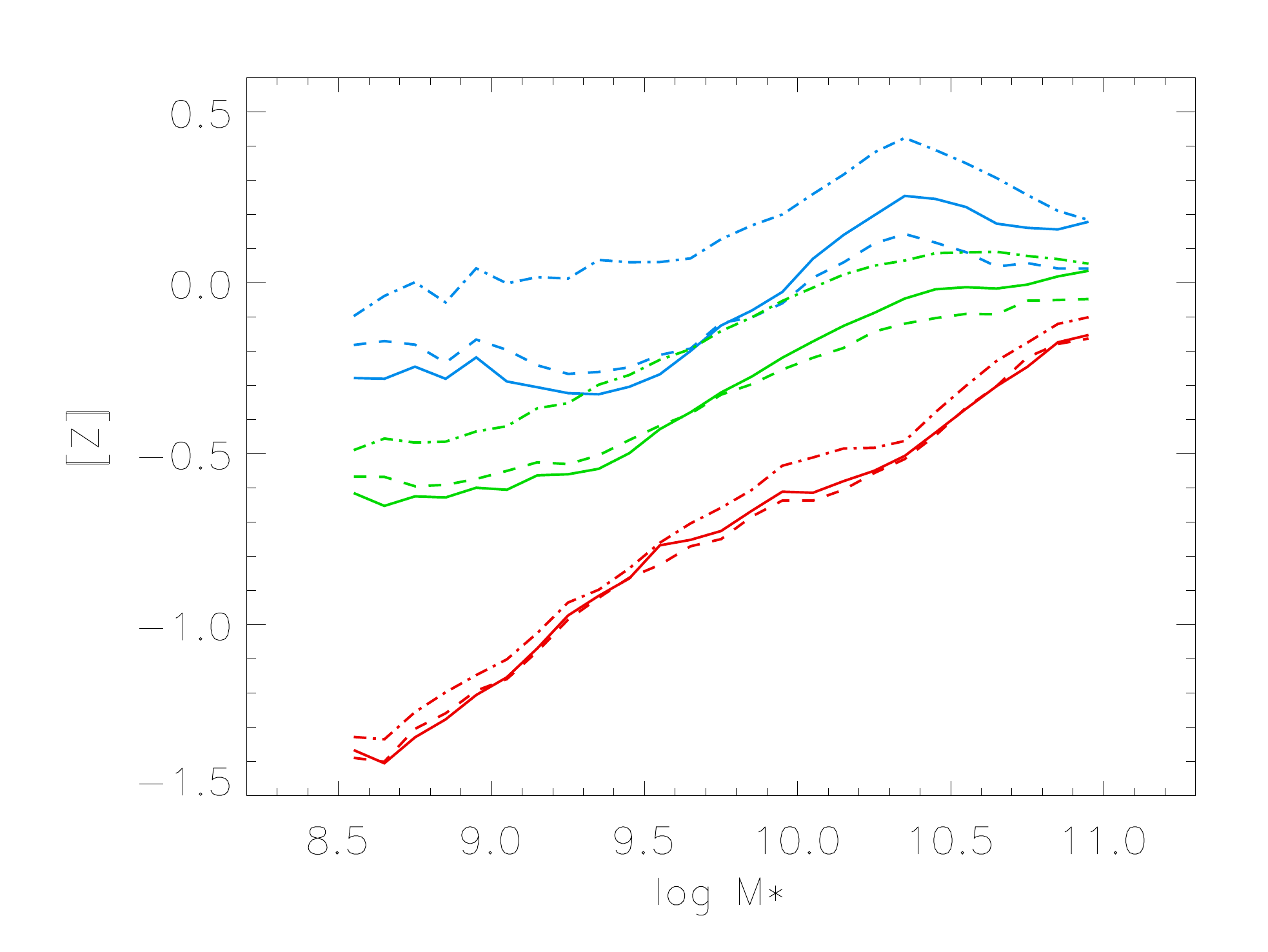}
    \includegraphics[scale=0.40]{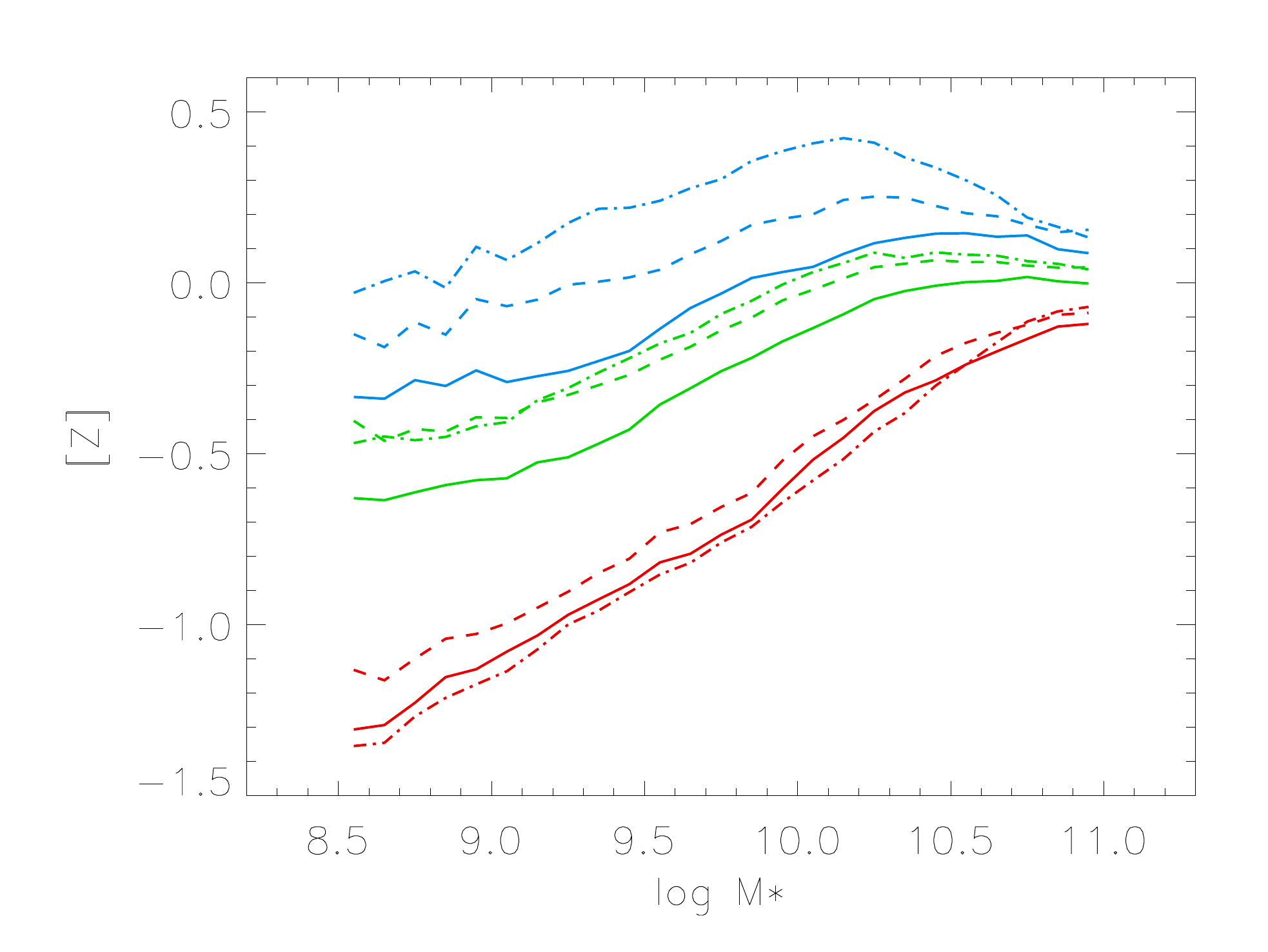}
    \includegraphics[scale=0.40]{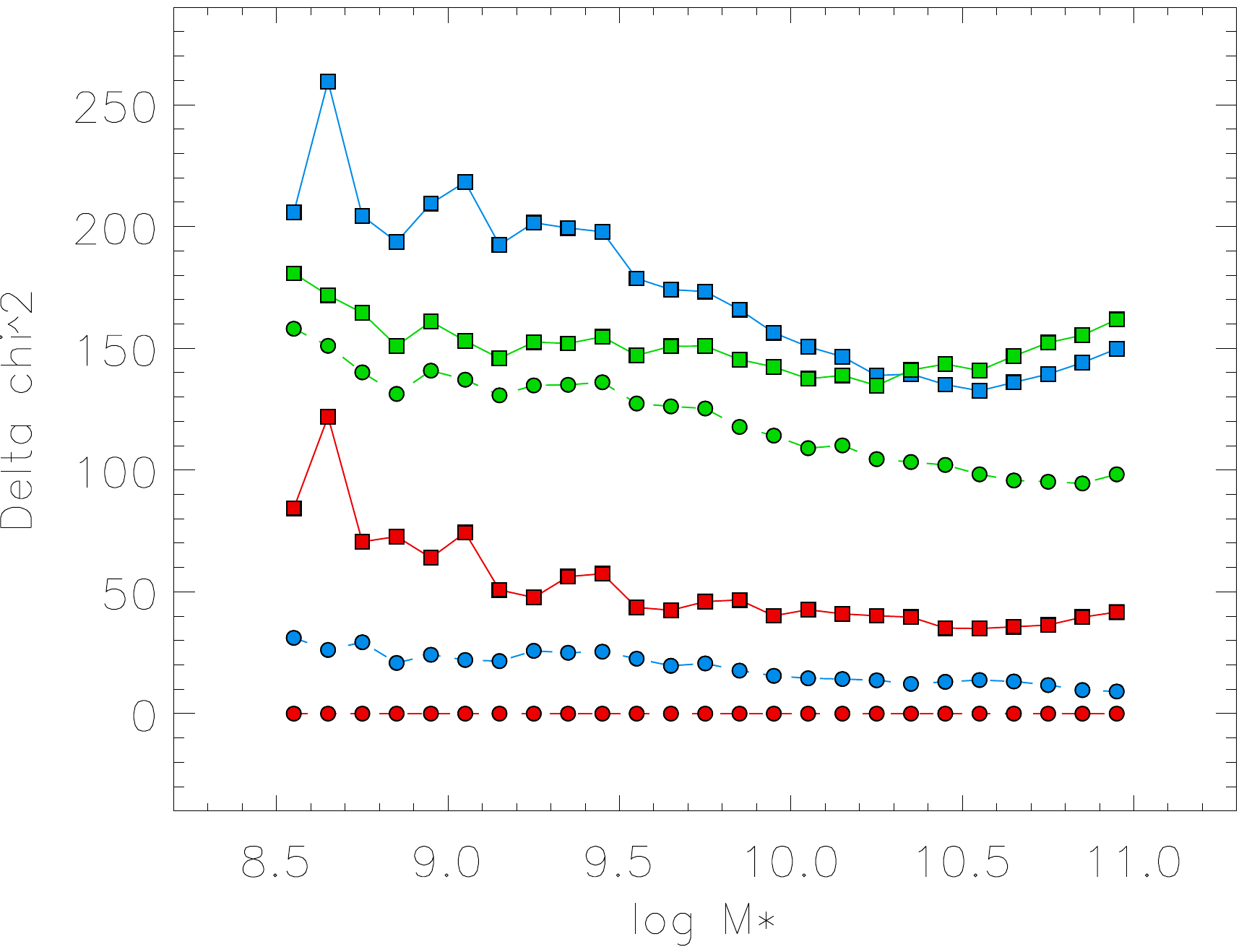}
  \end{center}
  \caption{Top: SDSS fit metallicities of Figure \ref{figure_7} (high time resolution) with all burst lengths overplotted (0.1 Myr: solid, 1 Myr: dashed, 10 Myr: dashed dotted). Middle: same as top but with metallicities of Figure \ref{figure_8} (low time resolution). Bottom: Assessement of the quality of the fits obtained with the different sets of SSB. Difference $\Delta \chi^2$ = $\chi^2_{min} - \chi^2_{min,0}$ for the different SSB spectral model fits as a function of galaxy stellar mass. $\chi^2_{min,0}$ is the value obtained with SSBs calculated with high time resolution and t$_b$ = 0.1 Myr. Circles correspond to SSB with high time resolution, squares to low resolution, red color to 0.1 Myr bursts, blue color to 1.0 Myr and green to 10 Myr. 
} \label{figure_9}
\end{figure}

\begin{figure}[ht!]
  \begin{center}
    \includegraphics[scale=0.47]{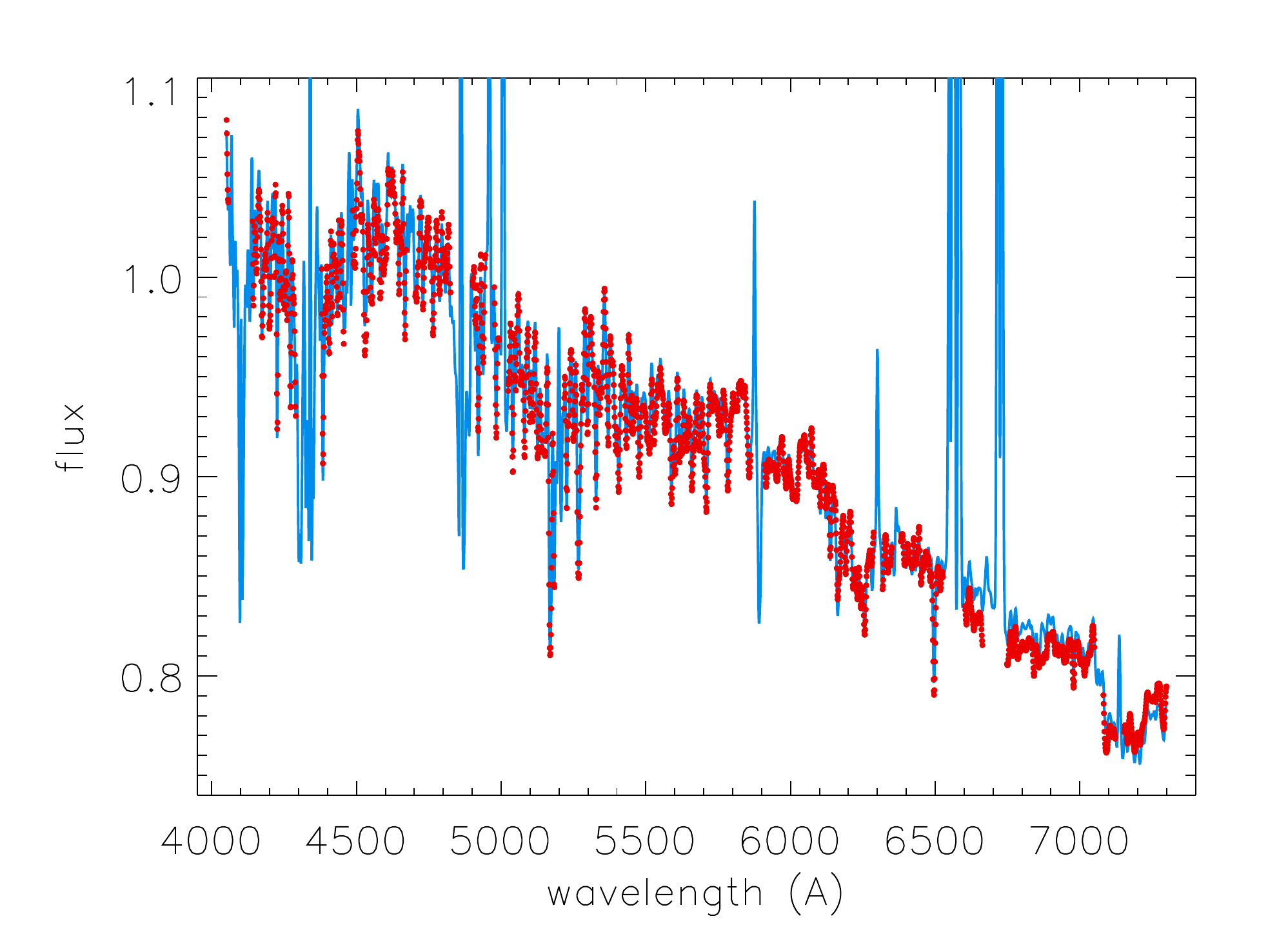}
  \end{center}
  \caption{
SSB spectral fit (red dots) of the observed spectrum (blue) for log M$_*$ = 9.75. Note that for the fit we mask out regions contaminated by ISM emission or absorption lines. The fit spectrum is calculated with a burst length t$_b$ = 0.1 Myr and high time resolution.
} \label{figure_10}
\end{figure}

\section{Analysis with Different Burst Lengths and Time Resolution}

In this section we use SSBs calculated for different burst lengths and time resolution to analyse our stacked SDSS spectra. We note that we allow for variations of R$_V$ in our spectral fits. As will be discussed in section 5, the R$_V$ obtained are substantially higher than 3.1 (see Figure \ref{figure_13}).

Figures \ref{figure_7} and \ref{figure_8} summarize the results with respect to the MZR. We show metallicities of the young and old population and average metallicities as a function of stellar mass obtained when using the different sets of SSBs and compare with with the MZR of massive supergiant stars, which has been obtained by quantitative spectroscopy of individual objects in 17 nearby galaxies (see \citealt{Bresolin2022} and references therein). These metallicities are accurate to 0.1 dex and serve as a benchmark for the young stellar population. The metallicity uncertainties of the SDSS spectral fits as shown in Figures \ref{figure_7} and \ref{figure_8} are up to 0.07 dex for the young and old population, respectively, and 0.03 dex for the average over all ages. They are obtained from Monte Carlo simulations with fits of the observed stacked spectra modified by adding Gaussian noise.

Figures \ref{figure_7} and \ref{figure_8} indicate that burst length and SSB time resolution have a significant effect of up to 0.4 dex on the determination [Z]$_{young}$, the metallicity of the young stellar population. The influence on the average metallicity [Z] is somewhat smaller (up to 0.2 dex) and the differences for the old population are below or about 0.1 dex and, thus, of the order of the fit uncertainties. A comparison of all metallicities obtained with different burst lengths is given in  Figure \ref{figure_9}.

The goodness of the fits is assessed in Figure \ref{figure_9} where we compare the $\chi^2$-values obtained with the different sets of SSB for the fit of the observed SDSS spectra. (We note that number of wavelength points used for the calculation of $\chi^2$ is n$_{pix}$ = 2523). Figure \ref{figure_9} shows that the SSB models with the shortest burst length of 0.1 Myr and high time resolution provide the best fit, although the high time resolution fits with 1 Myr burst length come close. It is reassuring that for these two cases the SDSS young stellar population metallicities are in very good agreement with the metallicities of our supergiant benchmark sample (see Figure \ref{figure_7}) . We take this as an indication that the population synthesis method is reliable.

Figure \ref{figure_10} shows the fit of the SDSS spectrum at log M$_*$  = 9.75 as a typical example.

\begin{figure}[ht!]
  \begin{center}
    \includegraphics[scale=0.47]{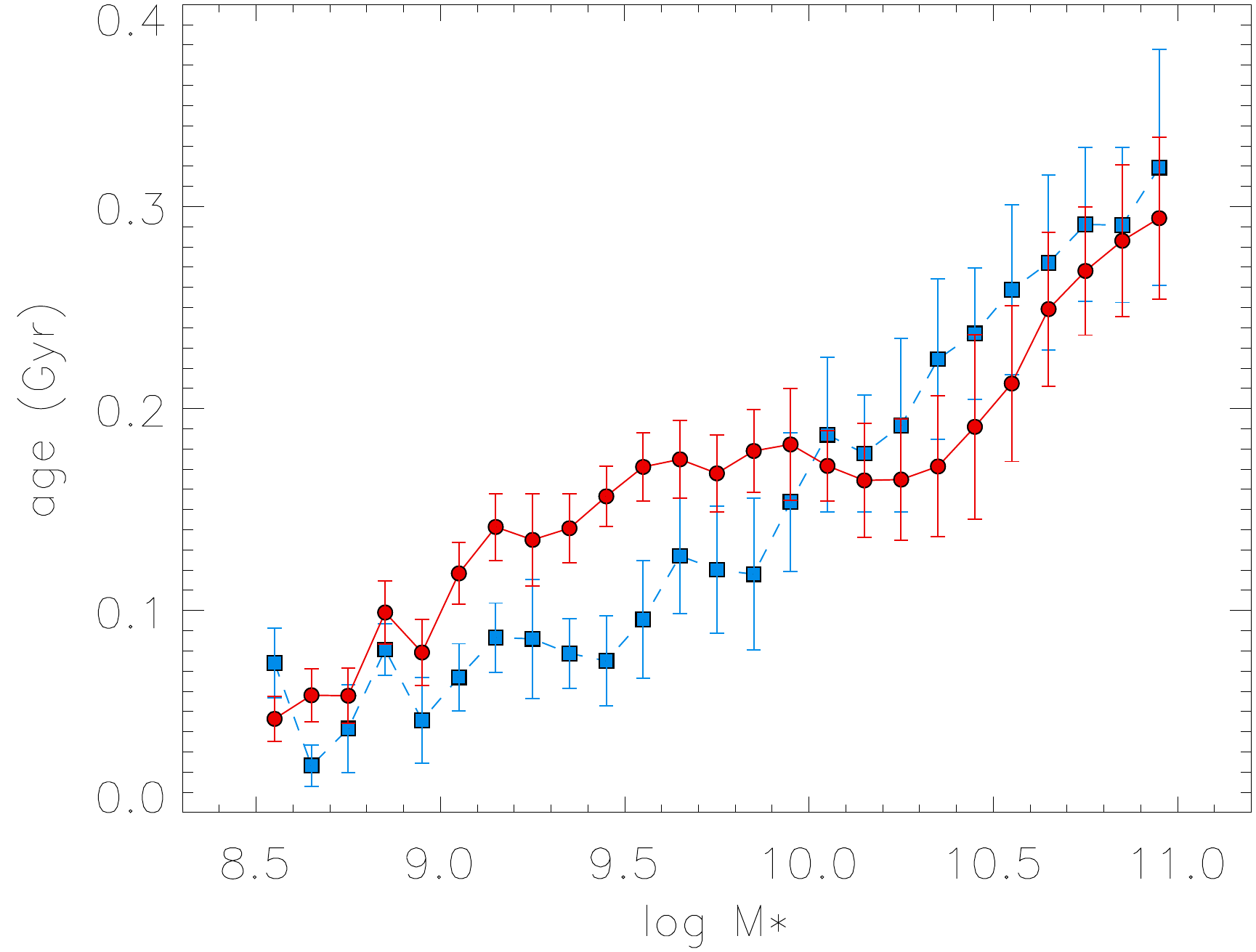}
    \includegraphics[scale=0.47]{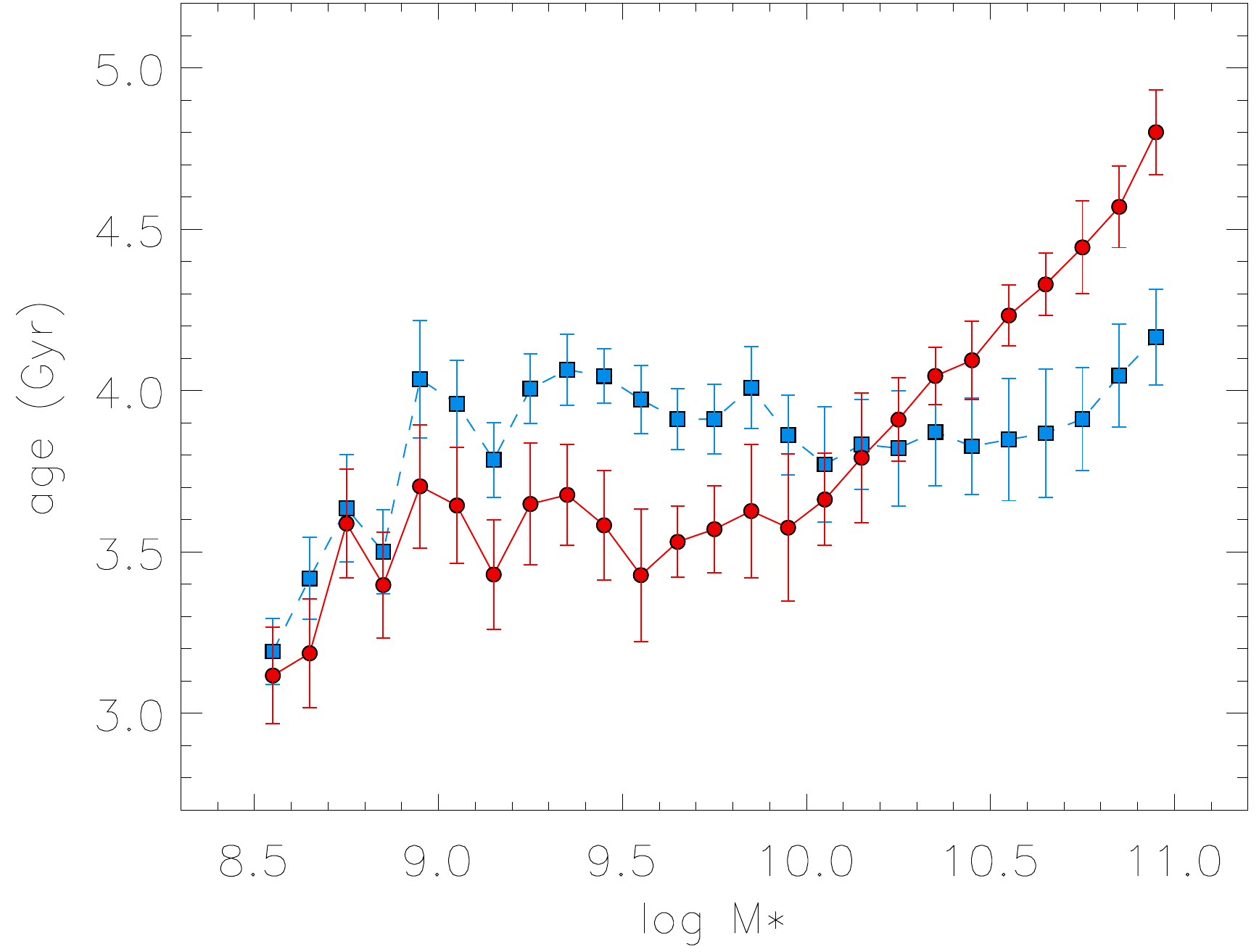}
    \includegraphics[scale=0.47]{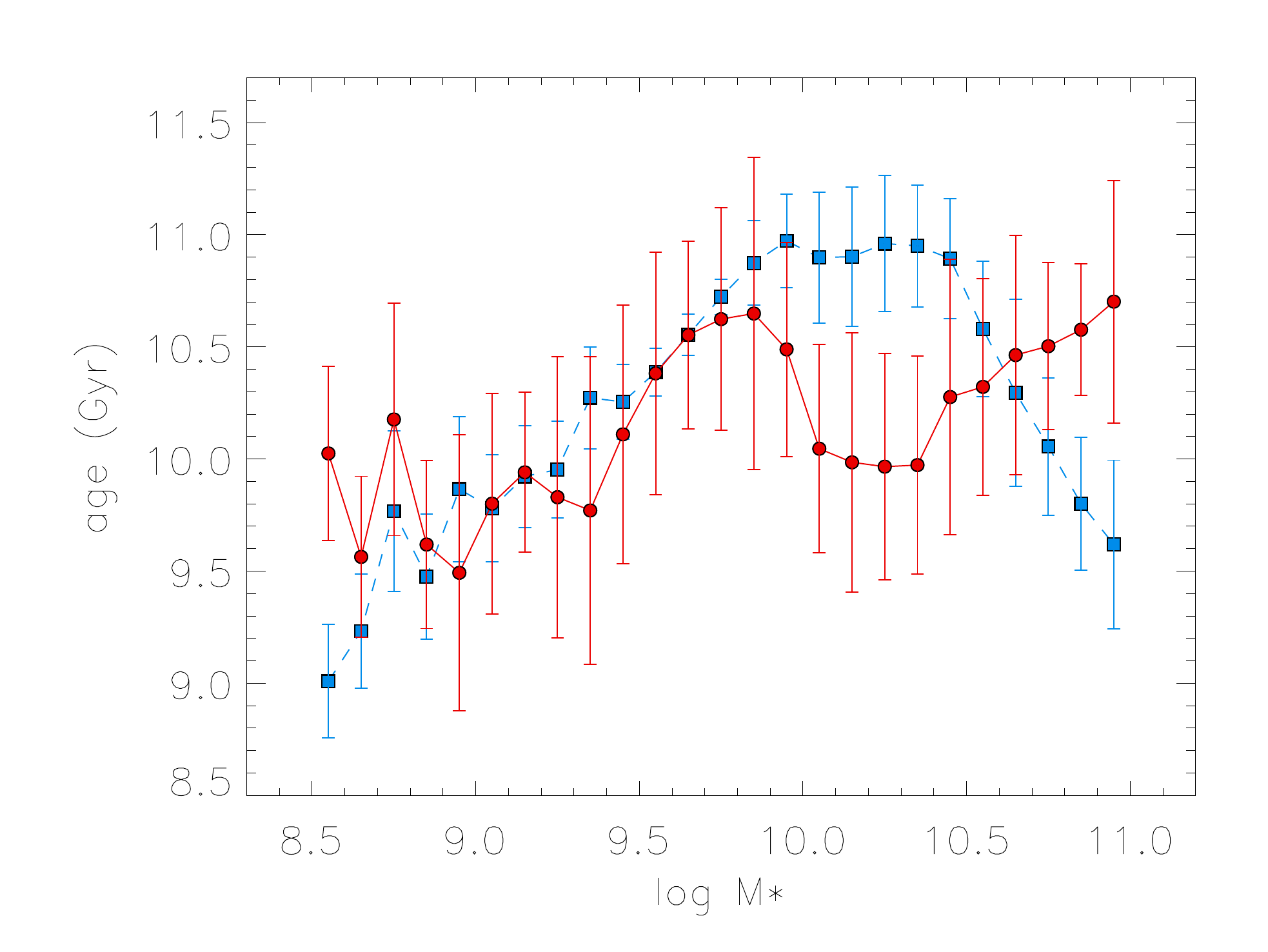}
  \end{center}
  \caption{
Ages t$_{young}$ (top),  t$_{av}$ (middle) and t$_{old}$ (bottom) as defined by equations (3), (7) and (8) as a function of galactic stellar mass obtaned with spectral fits suing high time reolution and a burst length t$_b$ = 0.1 Myr. Red values obtained with R$_V$ fitted. Blue values with R$_V$ = 3.1.
} \label{figure_11}
\end{figure}

\begin{figure}[ht!]
  \begin{center}
    \includegraphics[scale=0.47]{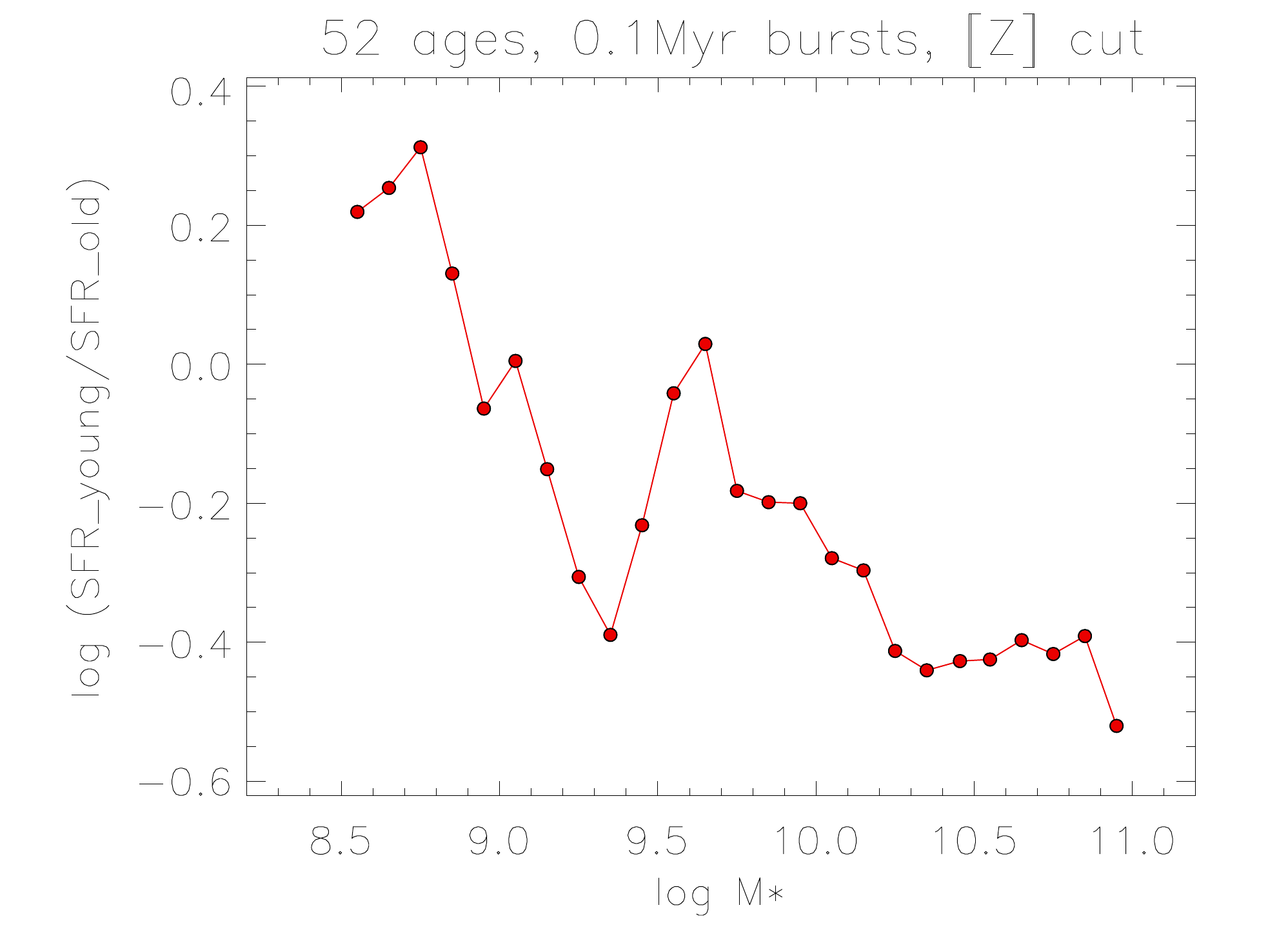}
  \end{center}
  \caption{
The ratio of star formation of young to the old stellar population as a function of galactic stellar mass. The fit of the observed spectra was carried out with t$_b$ = 0.1 Myr and high time resolution.
} \label{figure_12}
\end{figure}

The metallicities obtained for the old stellar population [Z]$_{old}$ are significantly lower than [Z]$_{young}$. This reflects galactic chemical evolution and will be discussed below.

We note that \cite{Zahid2017} in their work have used SSB with 10 Myr burst length and compared their average metallicities [Z] with the supergiant values finding good agreement. We confirm their result (see Figure \ref{figure_8}), but we note that this comparison is misleading. Since supergiants have ages around 50 Myr, a comparison with [Z]$_{young}$ would have been more appropriate.

\begin{figure}[ht!]
  \begin{center}
    \includegraphics[scale=0.47]{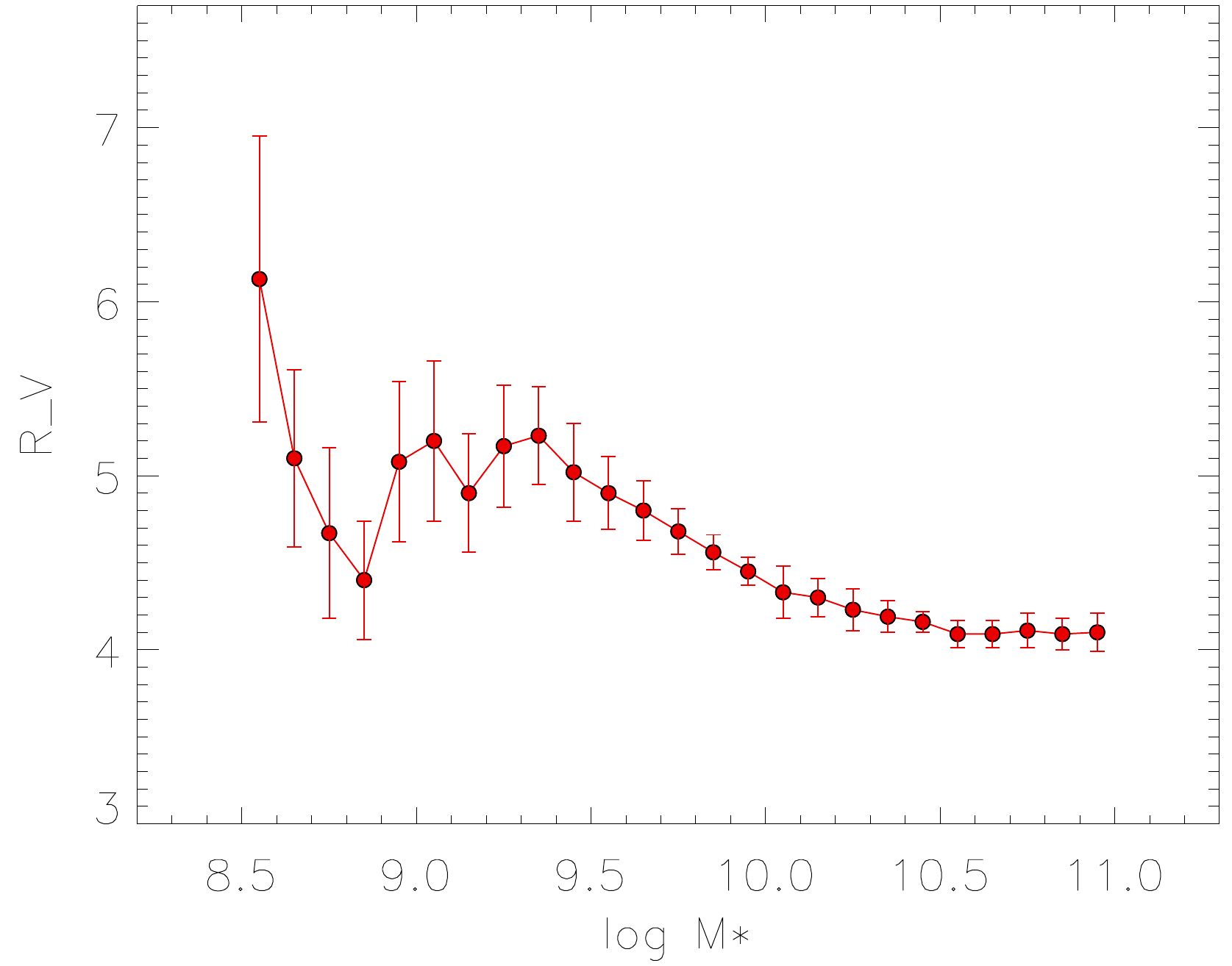}
    \includegraphics[scale=0.45]{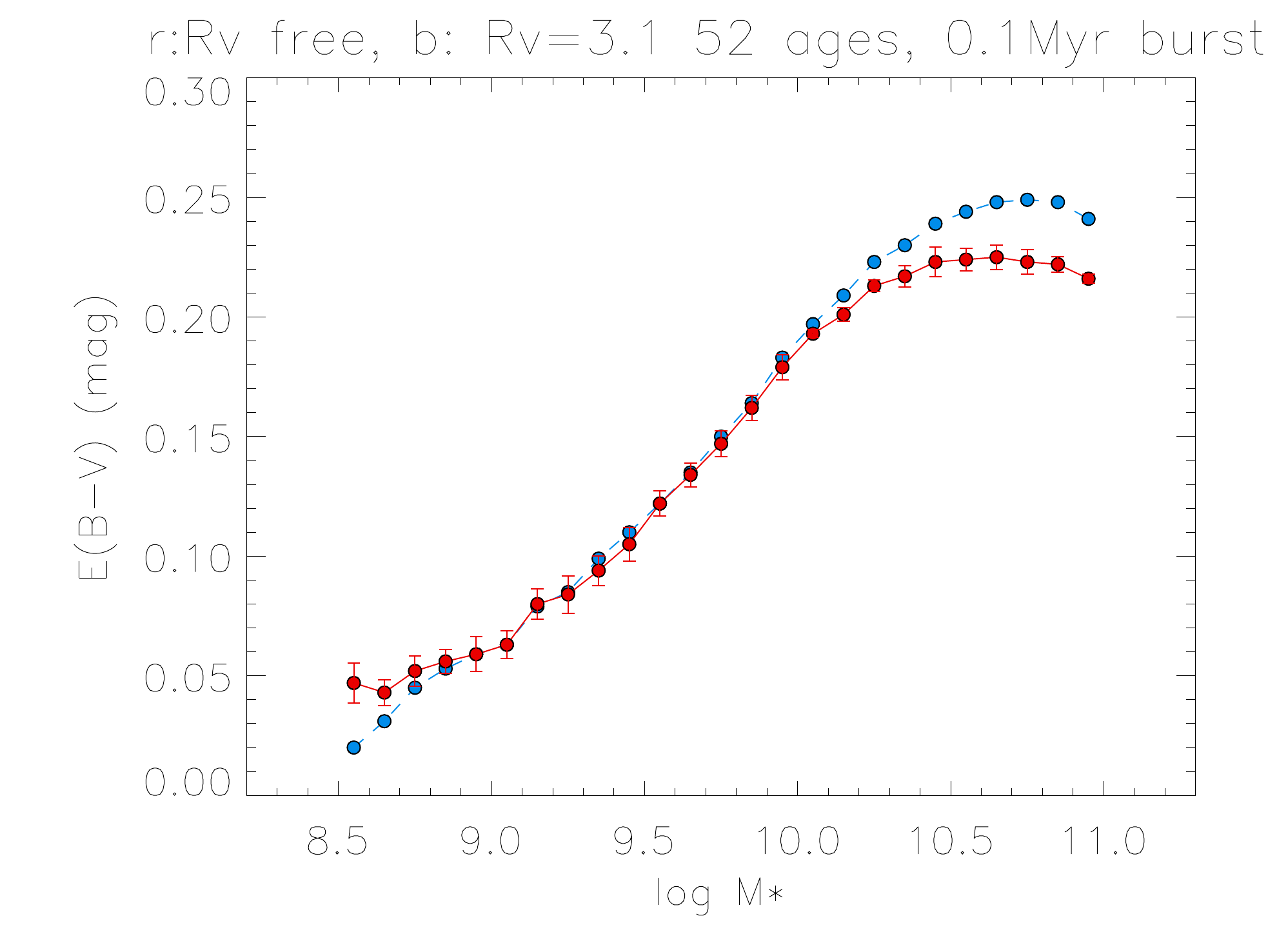}
    \includegraphics[scale=0.45]{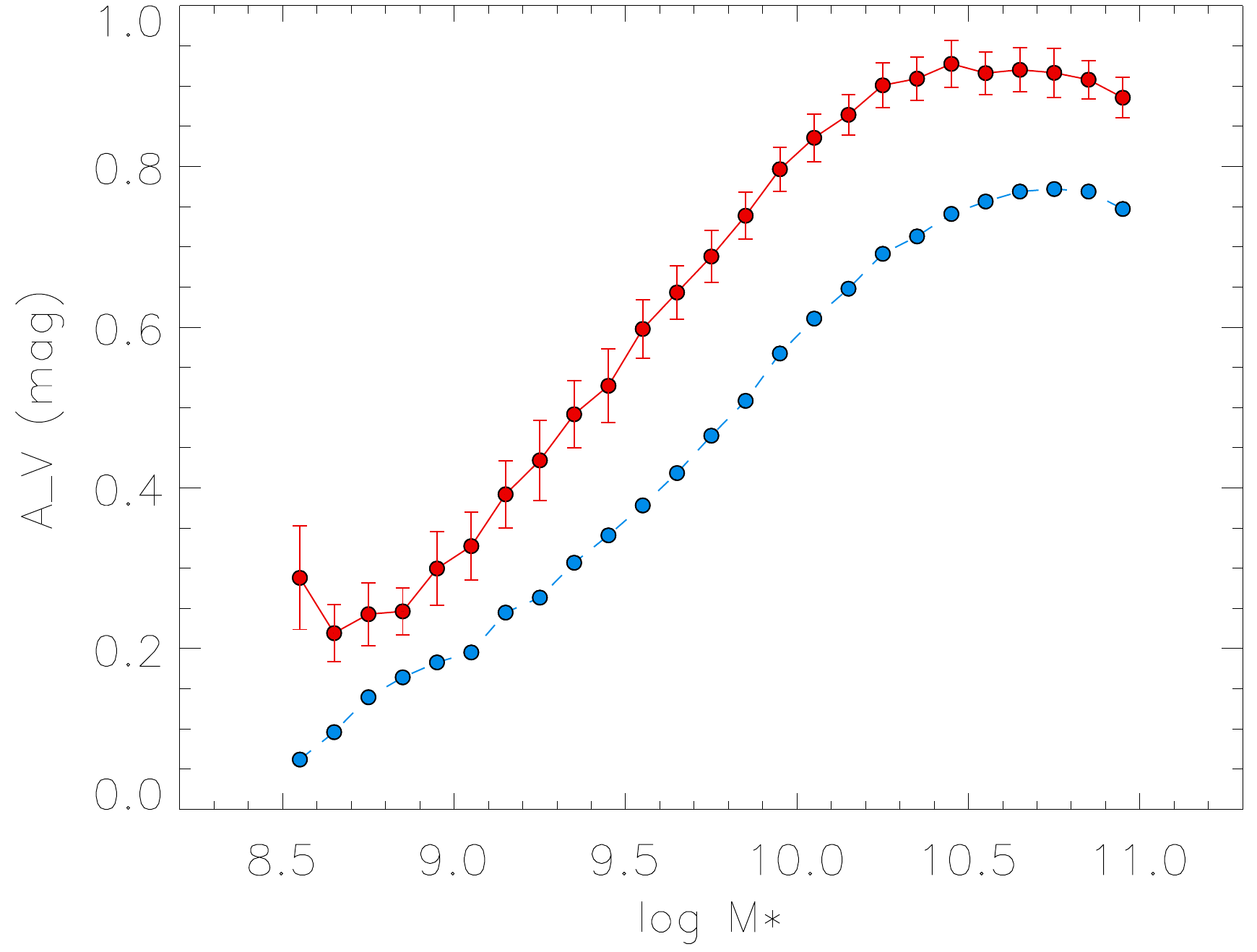}
  \end{center}
  \caption{
Top: The ratio of V-band extinction to reddening R$_V$ as a function of stellar mass log M$_*$. Middle: Reddening E(B-V) as a function of stellar mass. Red: values obtained with R$_V$ as a free parameter and determined as in the top figure. Blue: values obtained with R$_V$ = 3.1. Bottom: Interstellar V-band extinction A$_V$. Red and blue as in the middle figure. t$_b$ = 0.1Myr and high age resolution were used for the spectra fit.
} \label{figure_13}
\end{figure}

\begin{figure}[ht!]
  \begin{center}
    \includegraphics[scale=0.47]{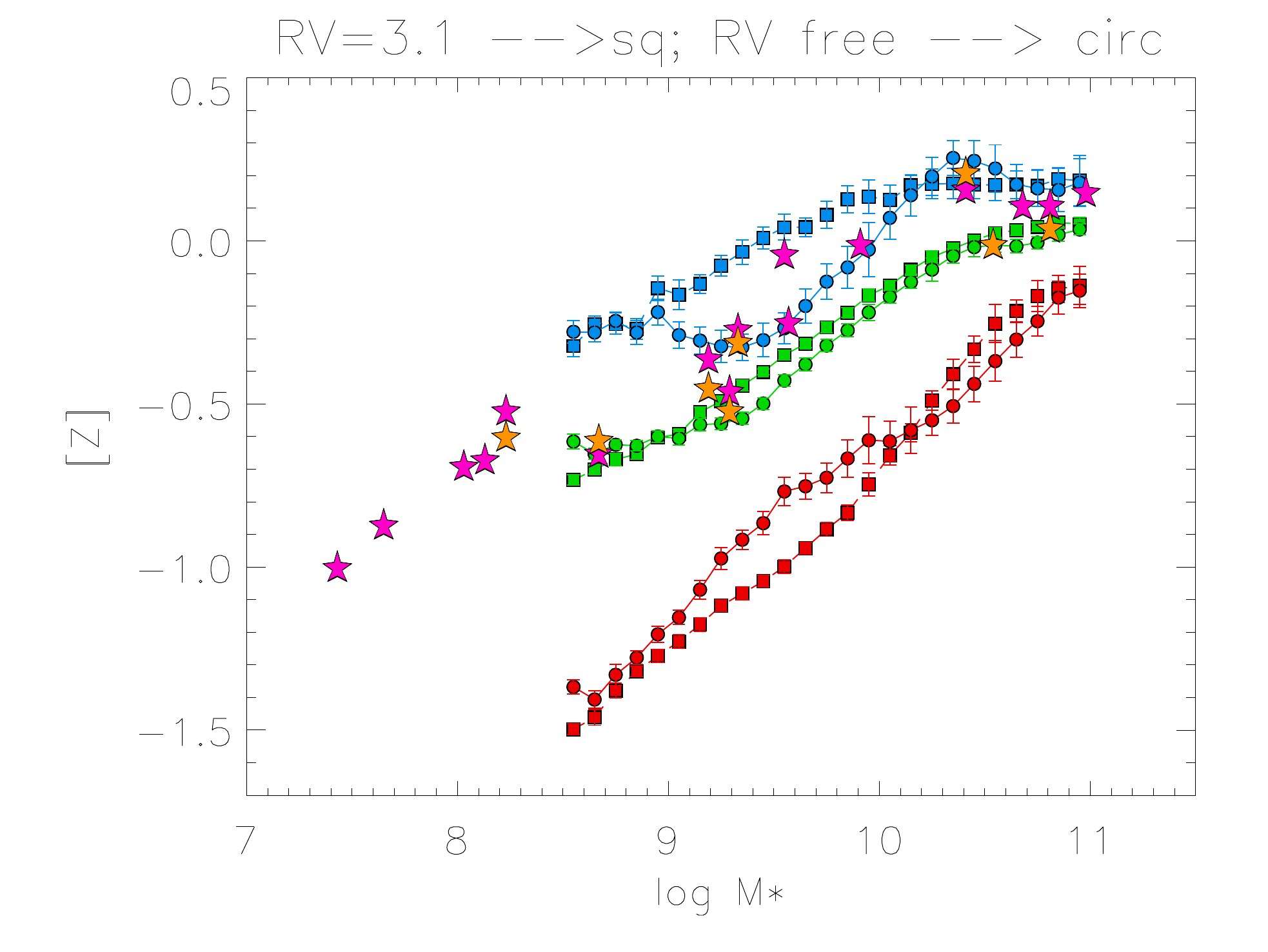}
  \end{center}
  \caption{
Same MZR as in Figure \ref{figure_7} (t$_b$ = 0.1Myr, high age resolution) but with the results of spectral fits with R$_V$ = 3.1 additionally shown as squares.
} \label{figure_14}
\end{figure}

The average stellar ages t$_{av}$ and the ages of the young and old population obtained with SSB of t$_b$ = 0.1 Myr and the high age resolution grid are displayed in Figure \ref{figure_11} together with their error bars. The ages between 50 and 300 Myr of the young population confirm that a comparison with supergiant stars as a benchmark is appropriate. We notice a clear correlation of t$_{young}$ with galactic stellar mass. No correlation is obtained for t$_{av}$ and the age t$_{old}$ of the old population.

\begin{figure}[ht!]
  \begin{center}
    \includegraphics[scale=0.46]{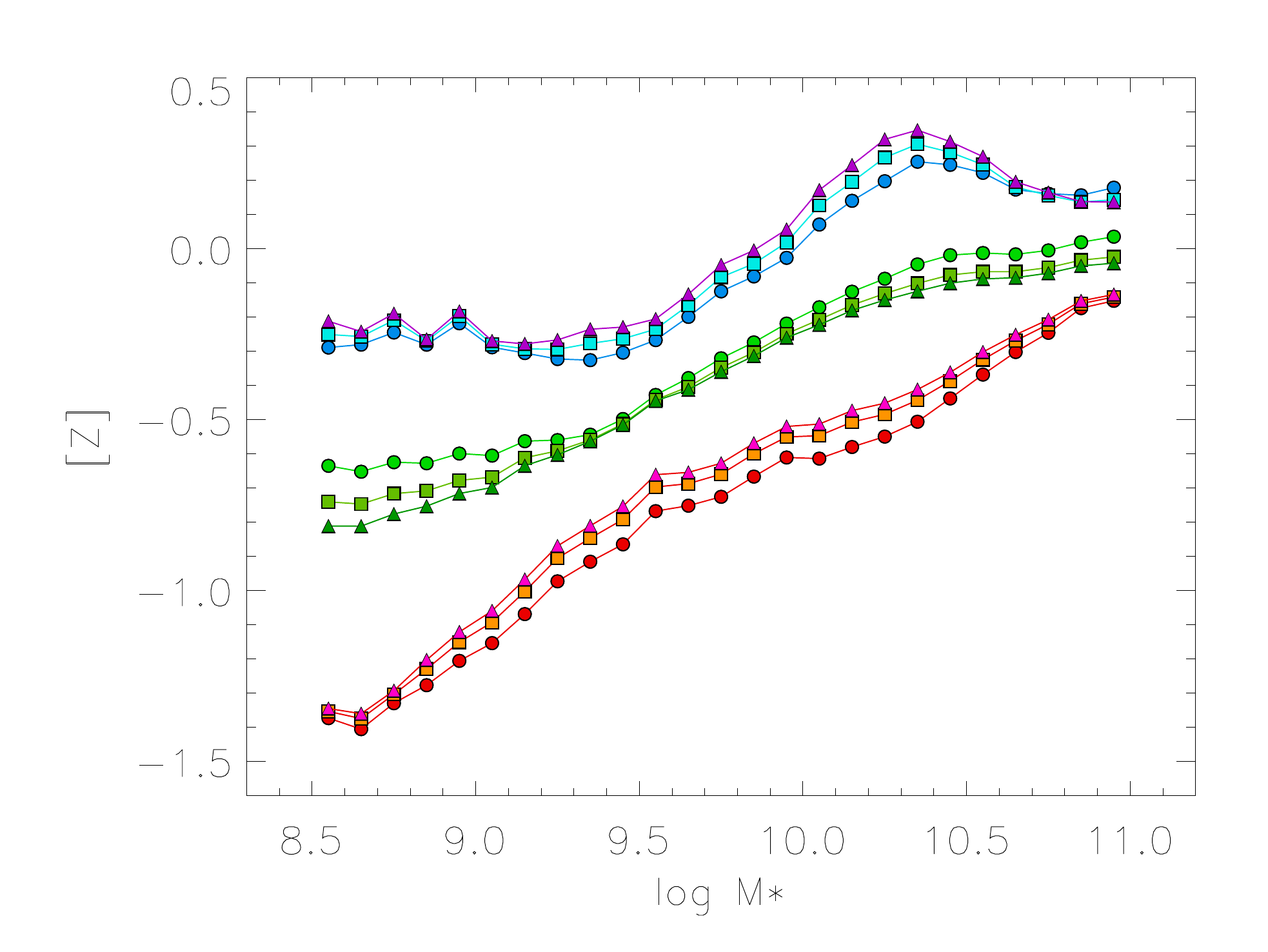}
    \includegraphics[scale=0.44]{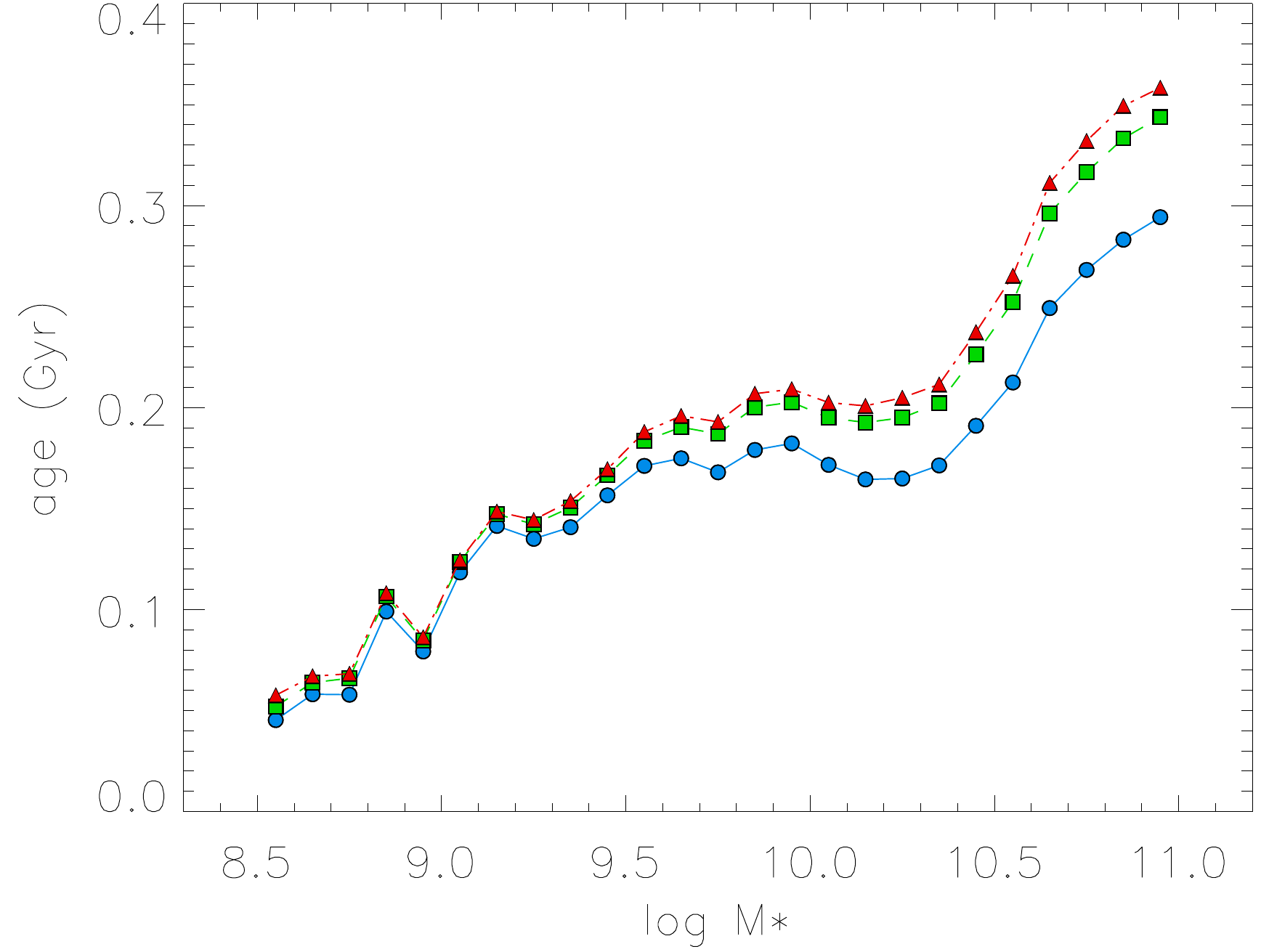}
    \includegraphics[scale=0.44]{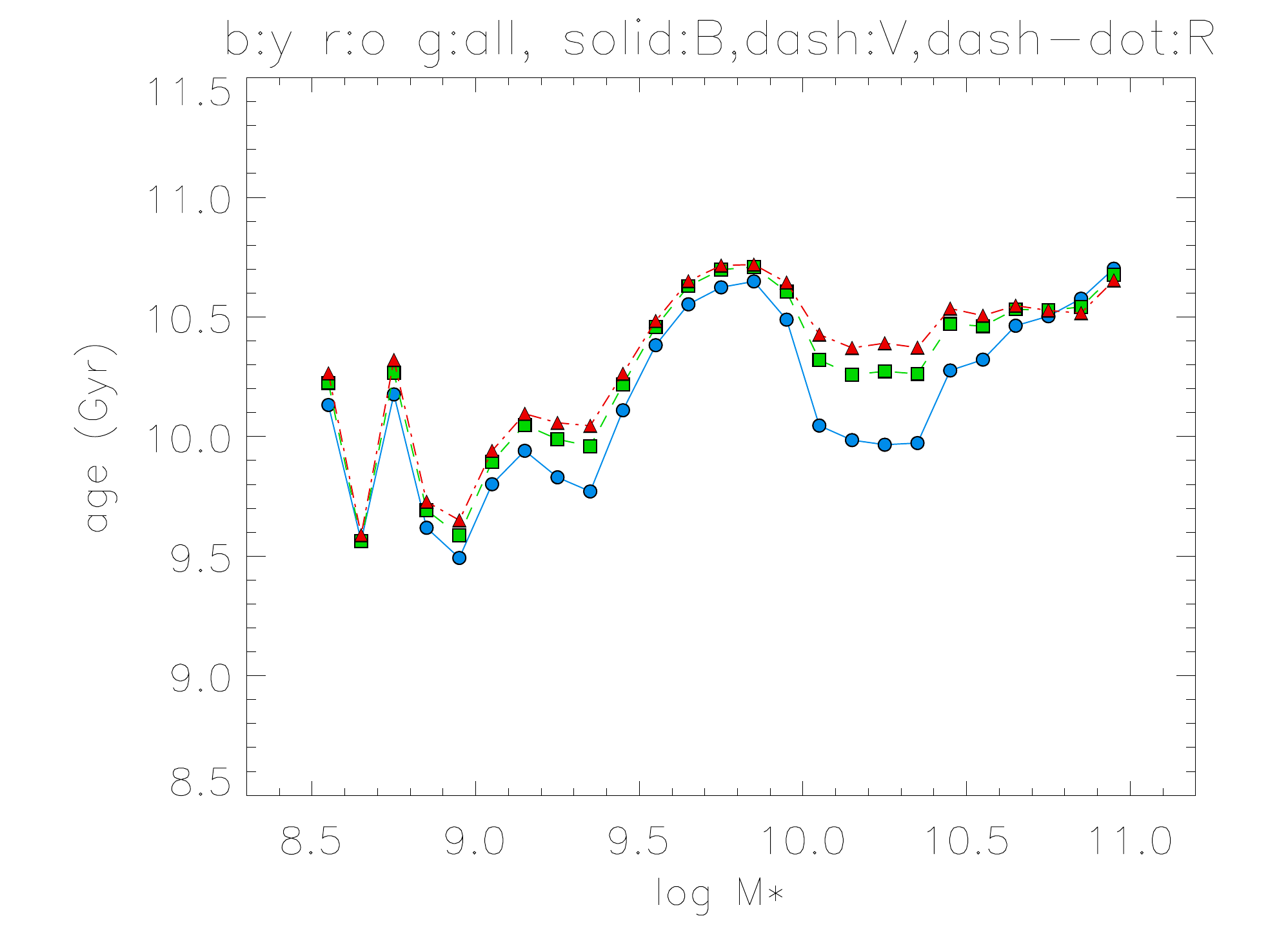}
  \end{center}
  \caption{
Top: Metallicities obtained with different normalization, where violet, cyan and blue belong to the young population, pink, orange and red to the old population and green colors show the average metallicities. Circles correspond to B-band normalization, squares to V-band, and triangles to R-band. Middle: Ages of the young population. Blue color: B-band normalization, green color: V-Band, red color: R-band. Bottom: Same as the middle plot, but for the old population. Again t$_b$ = 0.1Myr and high age resolution were used for the spectra fit.
} \label{figure_15}
\end{figure}

The average redshift of our galaxy sample is z=0.08 corresponding to a cosmological lookback time of 1.05 Gyr (assuming a Hubble constant H$_0$=70.4 km s$^{-1}$ Mpc$^{-1}$ and a flat universe with $\Omega_{\Lambda}$=0.728 and $\Omega_m$=1-$\Omega_{\Lambda}$). With t$_{old} \approx$ 10 Gyr we then know that the stars of the old population have been formed on average at a lookback time of 11 Gyr equivalent to a redshift of z $\approx$ 2.5. Thus, the metallicities obtained for the old stellar population in Figure \ref{figure_7} and \ref{figure_8}  correspond to z $\approx$ 2.5.

The fit of the observed spectrum yields the coefficients b$_i$ in eq. (1) for the individual bursts, which can be related to the relative number contribution N$_i$ of stars of isochrone i, for which the SSB model spectrum is calculated,  via

\begin{equation}
  b_i = N_i L_i(X).
\end{equation}

\begin{figure}[ht!]
  \begin{center}
    \includegraphics[scale=0.45]{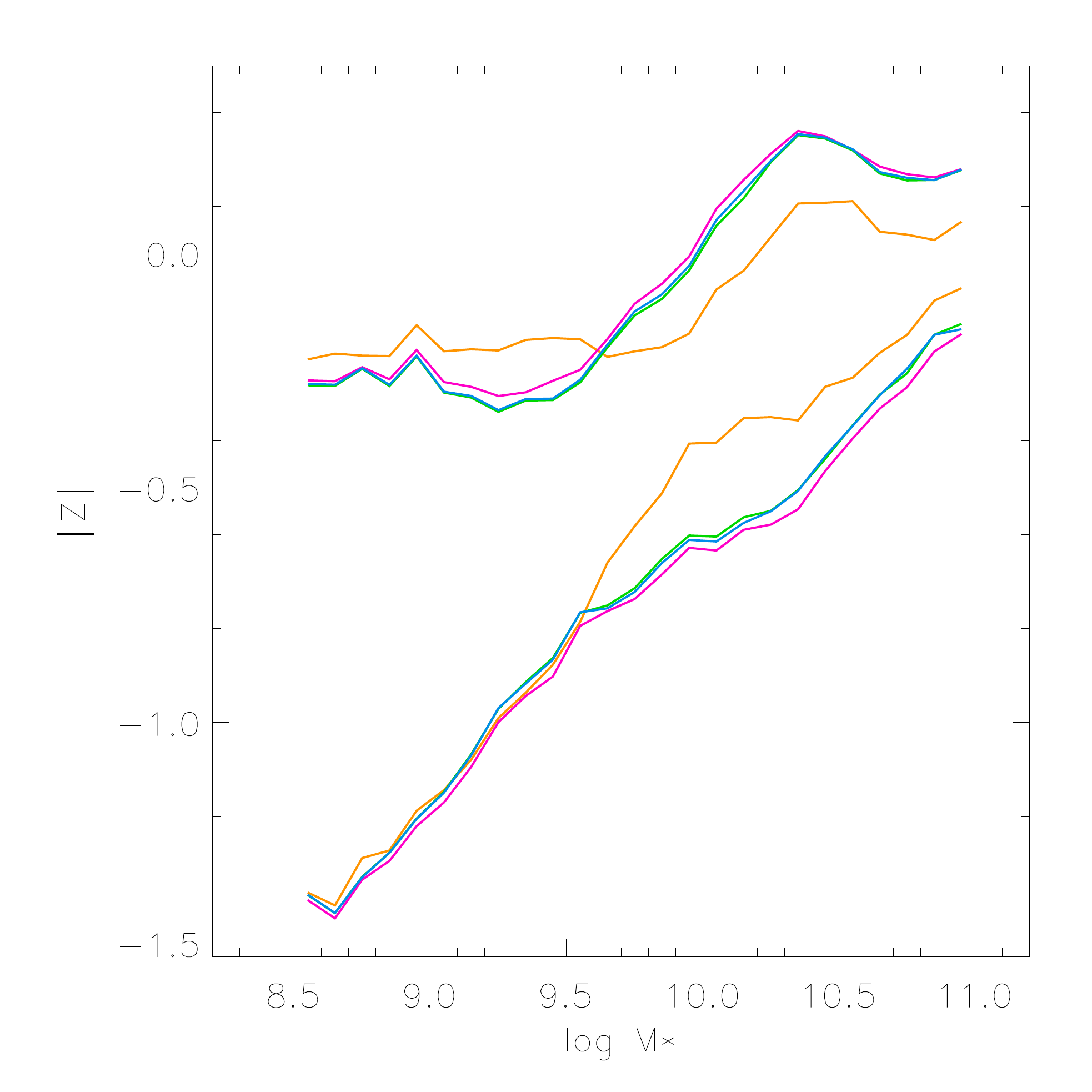}
  \end{center}
  \caption{
MZRs obtained by analysis with model spectra assuming different IMFs. Curves the in upper part of the figure correspond to the young population, curves in the lower part to the old population. Blue: \cite{Chabrier2003}, pink: \cite{Kroupa2001}, green: \cite{vanDokkum2008}, orange: \cite{Salpeter1955}.
} \label{figure_16}
\end{figure}

L$_i$(X) is the luminosity of the isochrone in passband X with an effective wavelength which corresponds to the wavelength at which the observed spectra and the SSB spectra are normalized. Since our normalization interval is 4400 to 4450 \angstrom\, the B-band is the appropriate choice for X (examples for  L$_i$(X) as a function of isochrone age are given in Figures \ref{figure_3} and \ref{figure_4}). Eq. (11) explains why the metallicities and ages of our spectral fits are B-band luminosity weighted quantities.  

Since N$_i\propto \psi_i \Delta$t$_i$, where $\Delta$t$_i$ is the time interval between isochrone i and the next older isochrone in our SSB grid, Eq. (11) can be used to estimate star formation rates $\psi$ for the young and old population

  \begin{equation}
    \psi_{young} \propto \sum_{i_{young}} {b_i \over \Delta t_i L_i}  ; \psi_{old}\propto \sum_{i_{old}} {b_i \over \Delta t_i L_i}. 
   \end{equation}

   Figure \ref{figure_12} shows the ratio of star formation of the young to the old population as a function of galaxy stellar mass. We see the obvious trend that low mass galaxies are presently more active than high mass galaxies.

\begin{figure}[ht!]
  \begin{center}
    \includegraphics[scale=0.47]{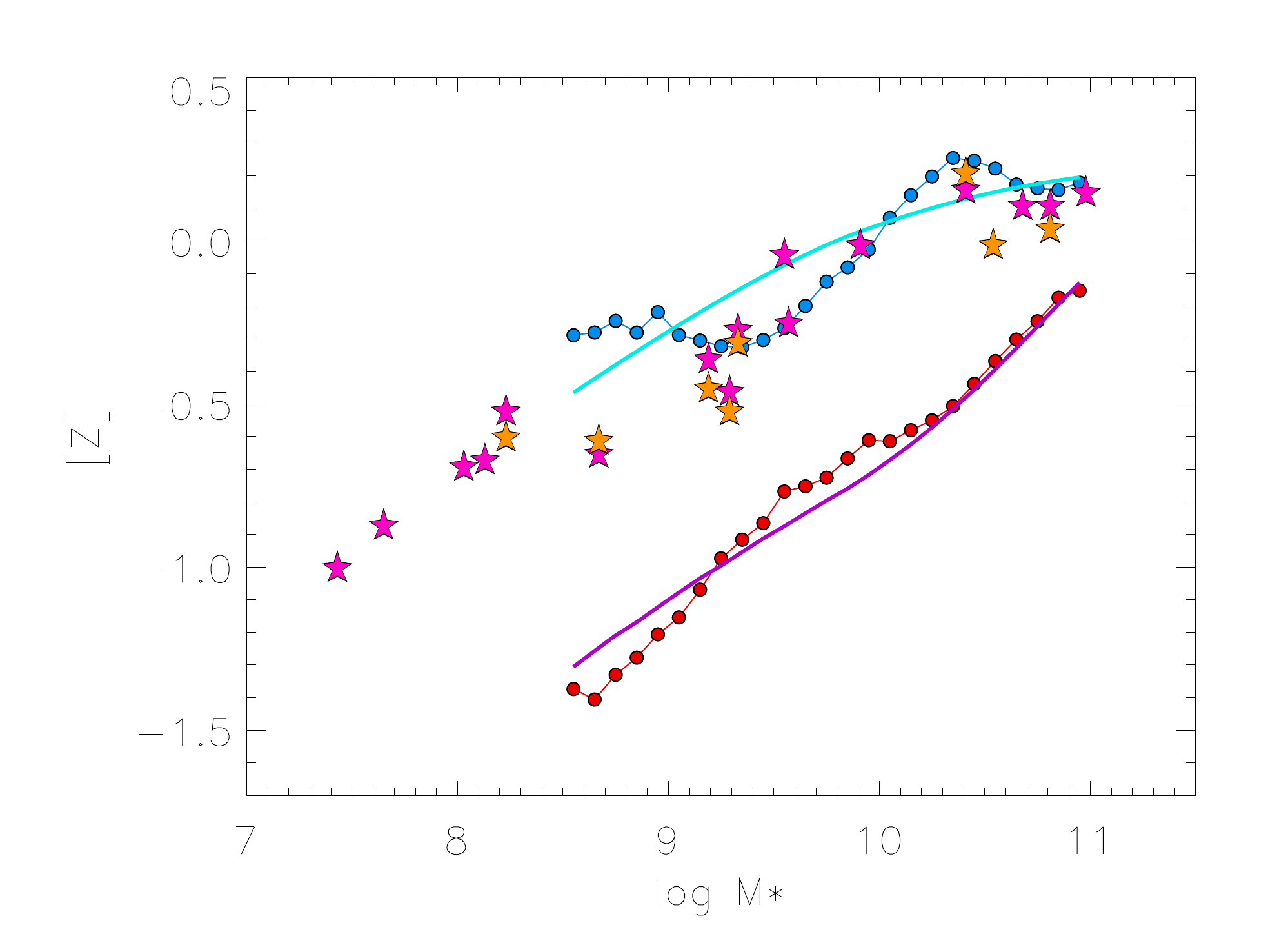}
    \includegraphics[scale=0.47]{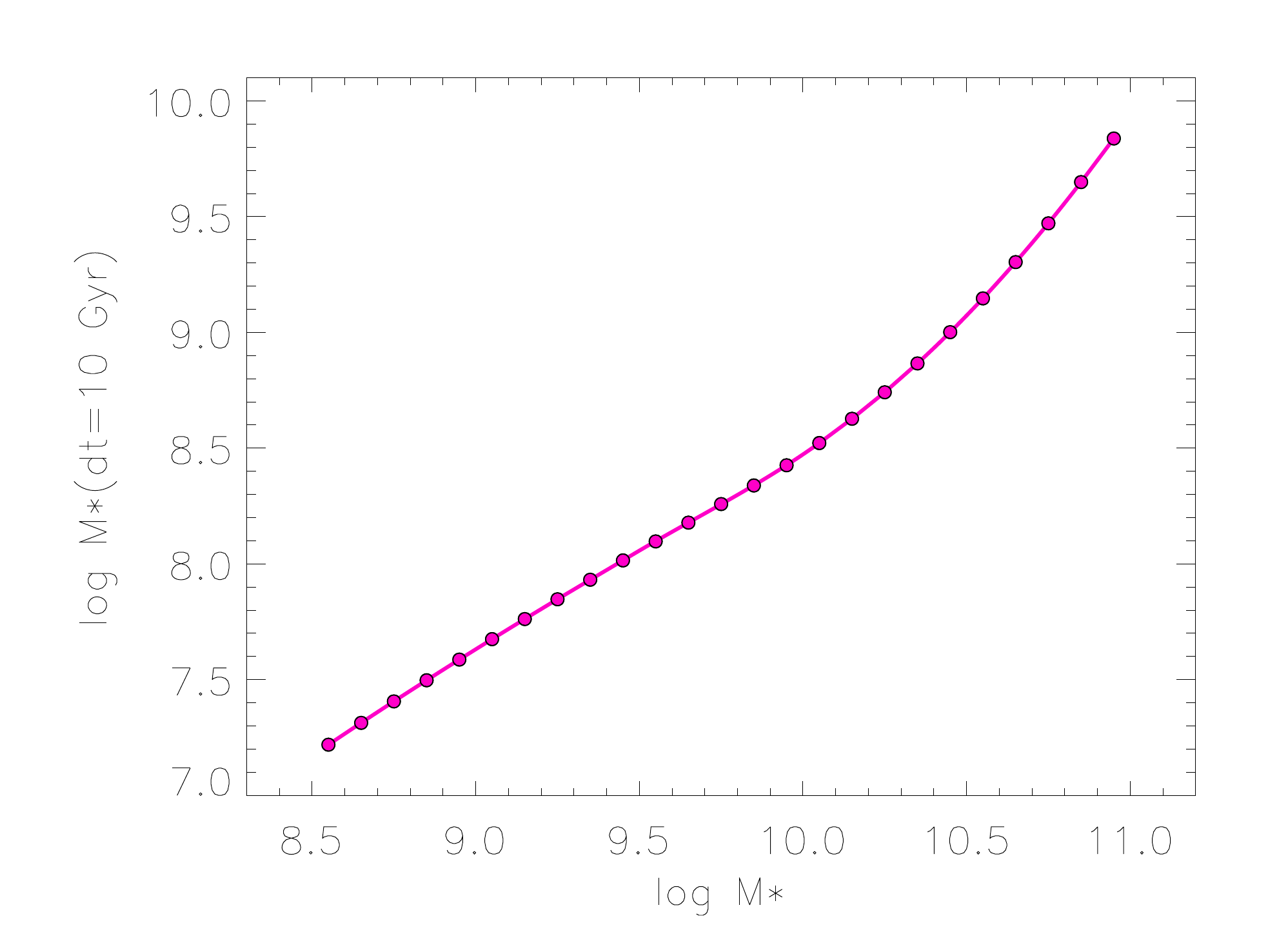}
  \end{center}
  \caption{
    The chemical evolution of star forming SDSS galaxies. Top: Metallicities of the young and old stellar population (as in Figure \ref{figure_7} top) and of supergiant stars compared with B-band luminosity weighted metallicities predicted by galaxy evolution models
    \citep{Kudritzki2021a, Kudritzki2021b} for very young stars (cyan) and stars 10 Gyr older (violet). Bottom: Mass of the galaxy evolution models at lookback time $\Delta$t = 10 Gyr as a function of galaxy final stellar mass.
} \label{figure_17}
\end{figure}
   
\section{Reddening and the Effects of the Extinction Law}

Since a long time there have been clear indications of deviations from the standard value R$_V$ = 3.1 in star forming regions of the Milky Way (see, for example, \citealt{Feinstein1973, Herbst1976, The1980, Cardelli1989}). Similar results have been found for the Large Magellanic Cloud, see \cite{Maiz2017}, \cite{urbaneja2017} and \cite{Holwerda2013} in their differential IFU SED study of two galaxy pairs conclude that the canonical extinction law with R$_V$ = 3.1 does not fit their data. In consequence, it is important to keep R$_V$ as free parameter for the SSB fits of the observed spectra as described in section 2. 

Figure \ref{figure_13} (top) shows the R$_V$ values obtained in our analysis for the case of t$_b$ = 0.1Myr and high age resolution. While the values at lower stellar masses have large uncertainties, we generally find values larger than 3.1 at all stellar masses. We believe the reason is that the SDSS observations sample the central regions of the star forming galaxies where still a lot of star formation in dense clouds is going on. These natal birthclouds act as an additional source of attenuation. In consequence, the contribution of the diffuse ISM with R$_V$ = 3.1 is small. 
We speculate that the reason for the obvious correlation with stellar mass is that the young stellar population is younger at lower mass (see Figure \ref{figure_11}) and that the density in the star forming regions is higher. 
The influence of varying density has been described by Witt \& Gordon in a series of seminal papers \citep{Witt1996, Gordon1997, Witt2000}. As a rule of thumb, an increasing inhomogeneity with alternating high- and low-density regions in the ISM structure leads to flattened ('grayer') attenuation curves. In our case, using a CCM/O'Donnell extinction law, a shallower slope is equivalent to high values of $R_V$. See also \cite{Berlind1997} for an in depth discussion of this issue in case of the galaxy NGC2207.

We note that the average value found for the LMC (log M$_*$ = 9.2) by \cite{urbaneja2017}, R$_V$ = 4.6, is in good agreement with Figure \ref{figure_13}. It is important to note that the difference from the standard reddening law in star forming galaxies at all masses has important consequences for other aspects of astrophysics. For instance, the estimate of galaxy luminosities and the corresponding mass to light ratios can be affected. Distance determinations based on stellar distance indicators and the resulting cosmological distance ladder might be influenced as well (see, for instance, \citealt{urbaneja2017,Falco1999}). 

Reddening E(B-V) and V-band extinction A$_V$ are also shown in Figure \ref{figure_13} (middle and bottom, respectively). Again, there is a clear correlation with stellar mass. Such correlations have already been found in previous work \citep{Brinchmann2004, Asari2007, Garn2010, Zahid2013b}. We note, however, that the deviation of R$_V$ from the standard value 3.1 leads to significantly higher extinction of about 0.2 magnitudes.

Ignoring a potential variation of R$_V$ and restricting to values of 3.1 also affects the determination of metallicities and the MZR. This is demonstrated in Figure \ref{figure_14}, where results obtained with R$_V$ = 3.1 are over-plotted. With differences up to 0.4 dex the effects are significant for the young and old population, most importantly in the middle mass range.

The determination of ages t$_{young}$, t$_{old}$, and t$_{av}$ is also affected, if a fixed value R$_V$ = 3.1 is adopted (see Figure \ref{figure_11}). The differences can be as large 50\%  and 10\% for the young and population, respectively, and 15\% for the average stellar age.

\begin{figure}[ht!]
  \begin{center}
    \includegraphics[scale=0.47]{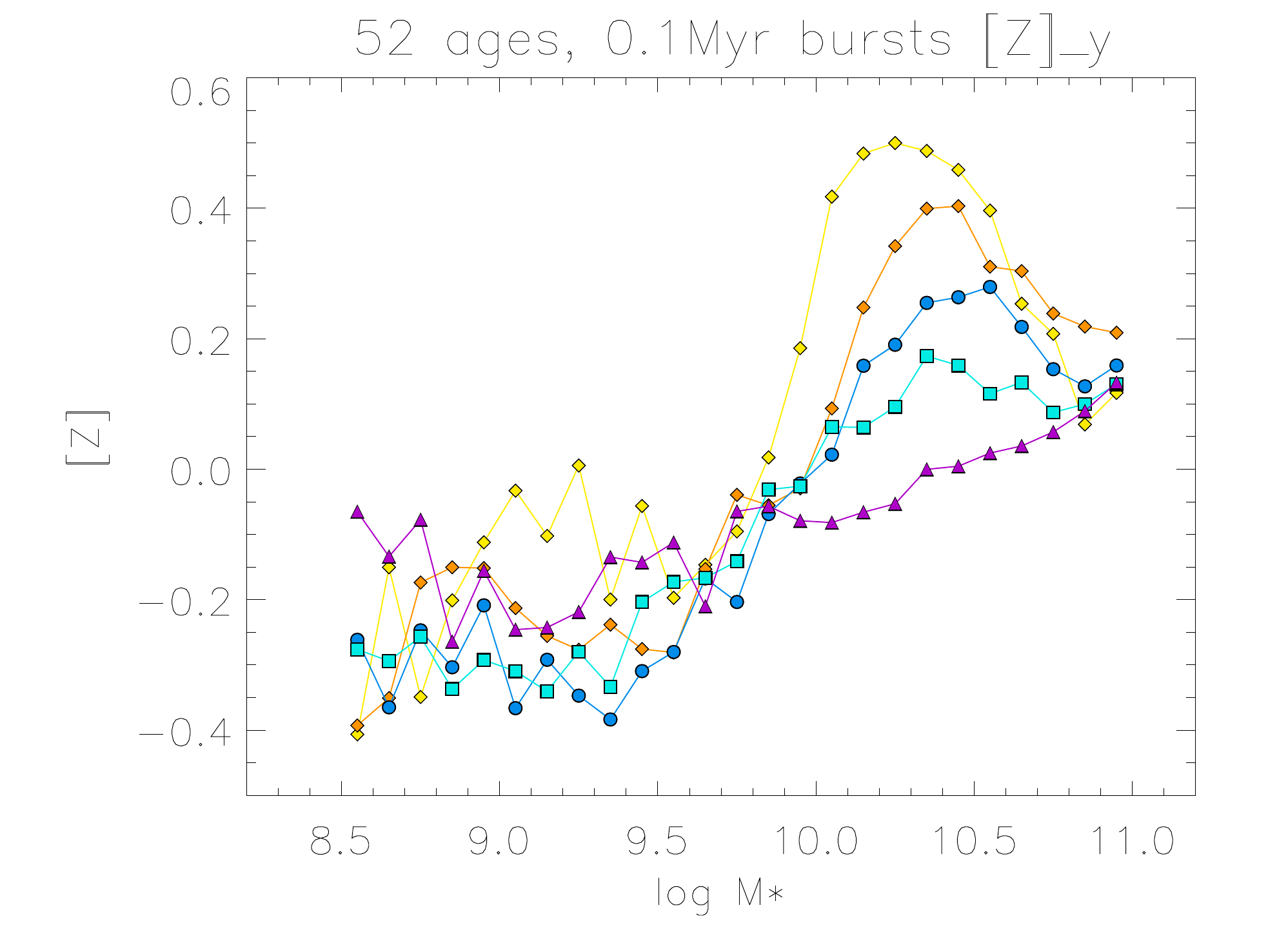}
    \includegraphics[scale=0.47]{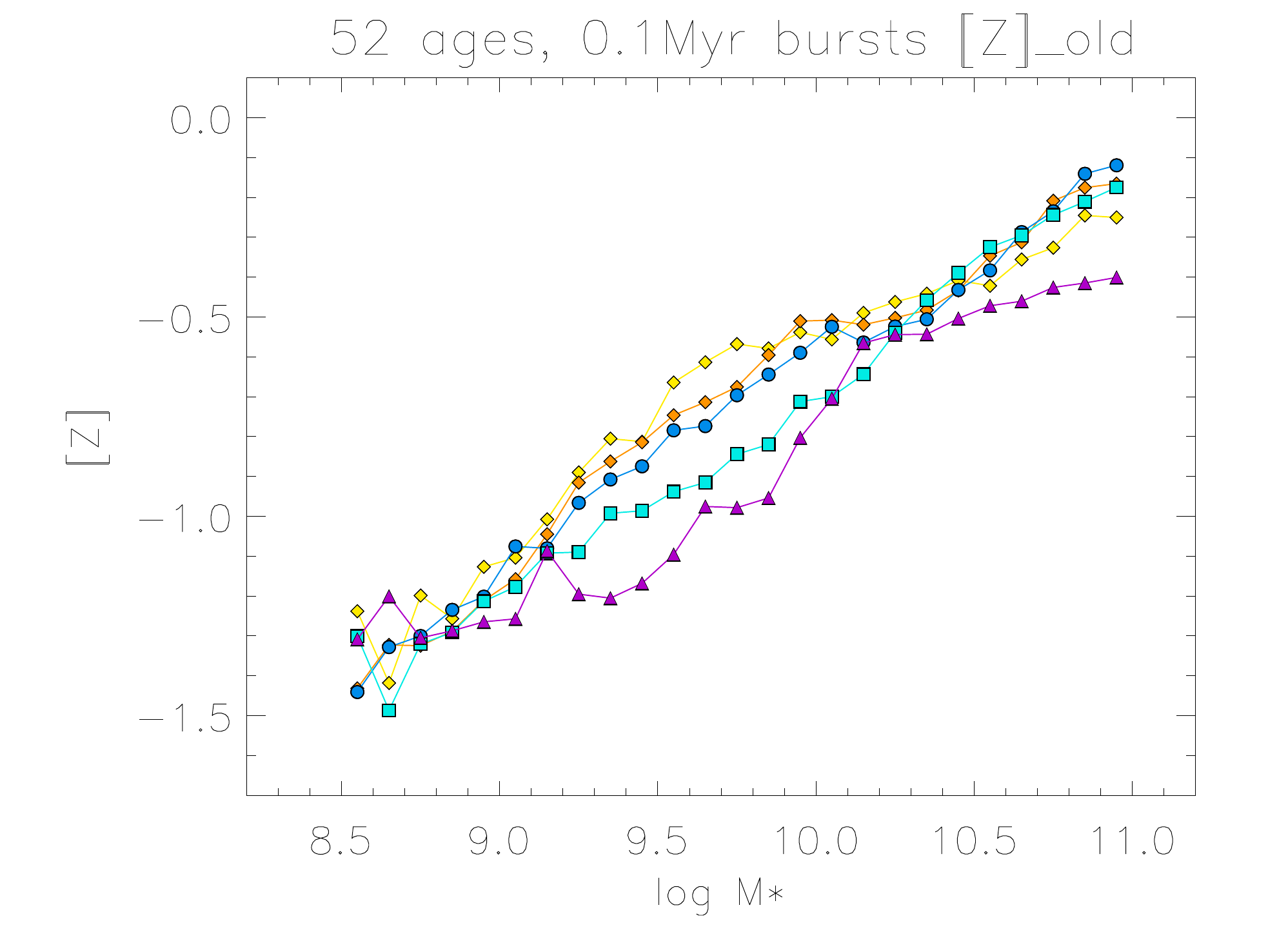}
  \end{center}
  \caption{
Metallicities of star forming SDSS galaxies stacked in quintiles of star formation rates at each stellar mass. Yellow, orange, blue, cyan, violet colors correspond to the different quintiles from lowest to highest SFR. Top: MZR of the young population, bottom: old population. 
} \label{figure_18}
\end{figure}

\section{Spectral Normalization}

The metallicities and ages obtained in the previous sections are B-band luminosity averages because we have normalized our spectra by setting the mean flux between 4400 and 4450 \angstrom\ to unity. This is illustrated by eq. (11).  It is, therefore, important to investigate how metallicities and ages change depending on the wavelength interval selected for normalization. For that purpose we have repeated the analysis with normalizations between 5500 and 5550 \angstrom\ (V-band) and 6950 and 7000 \angstrom\ (R-band). The comparison of the metallicities and ages obtained with the different normalizations are shown in Figure \ref{figure_15}.

The differences between the different normalizations are small. The metallicities of the young and old population  increase gradually with the normalization shifted towards the red, but the maximum difference is $\sim$ 0.1 dex only slightly larger than the metallicity uncertainties discussed in section 4. Ages also increase by a small amount but the differences are comparable to the errors shown in Figure \ref{figure_16}.

We note that R$_V$, E(B-V), A$_V$ and the star formation histories (as displayed in Figure \ref{figure_12}) remain unchanged and do not depend on the spectral normalization interval.

\begin{figure}[ht!]
  \begin{center}
    \includegraphics[scale=0.47]{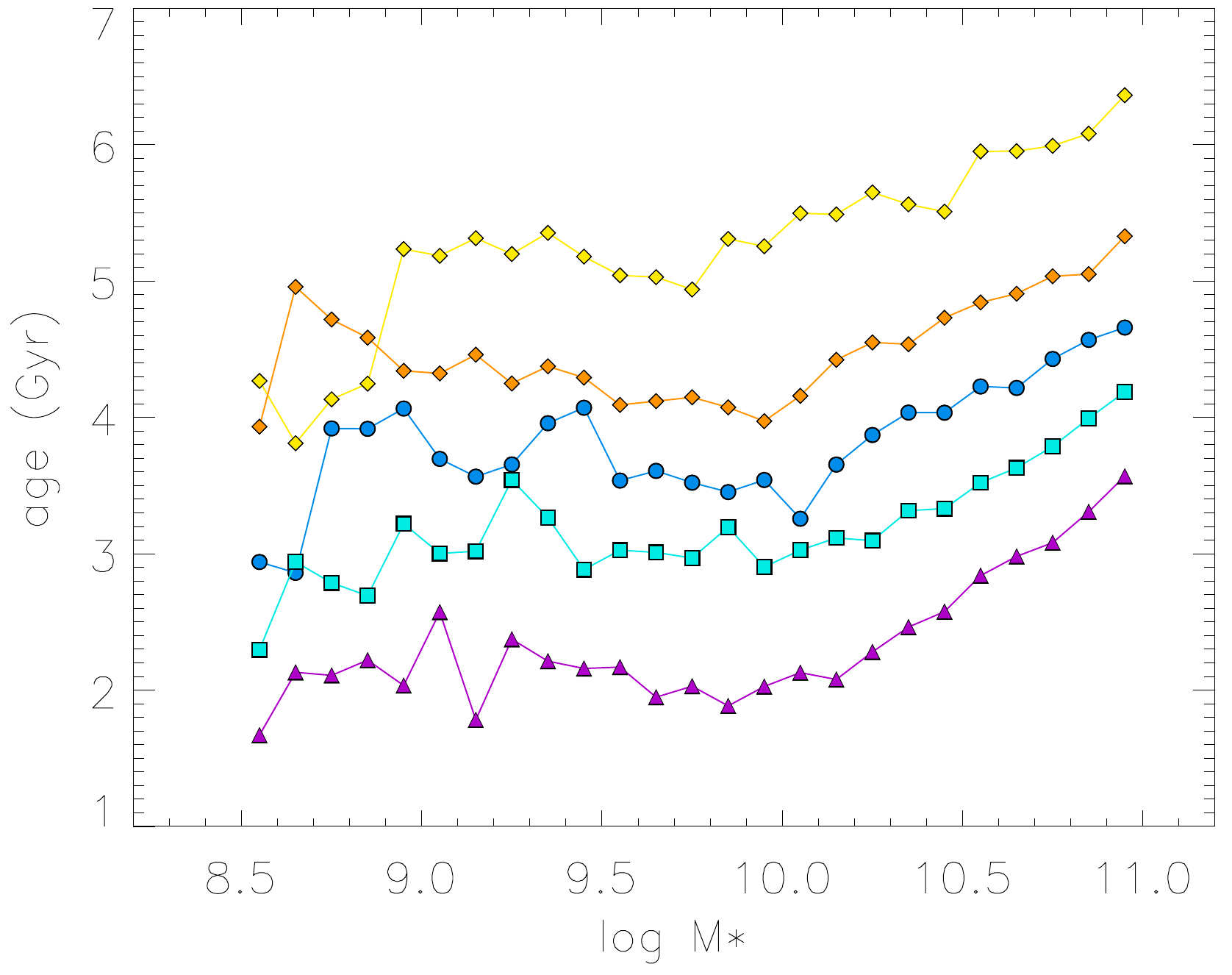}
    \includegraphics[scale=0.47]{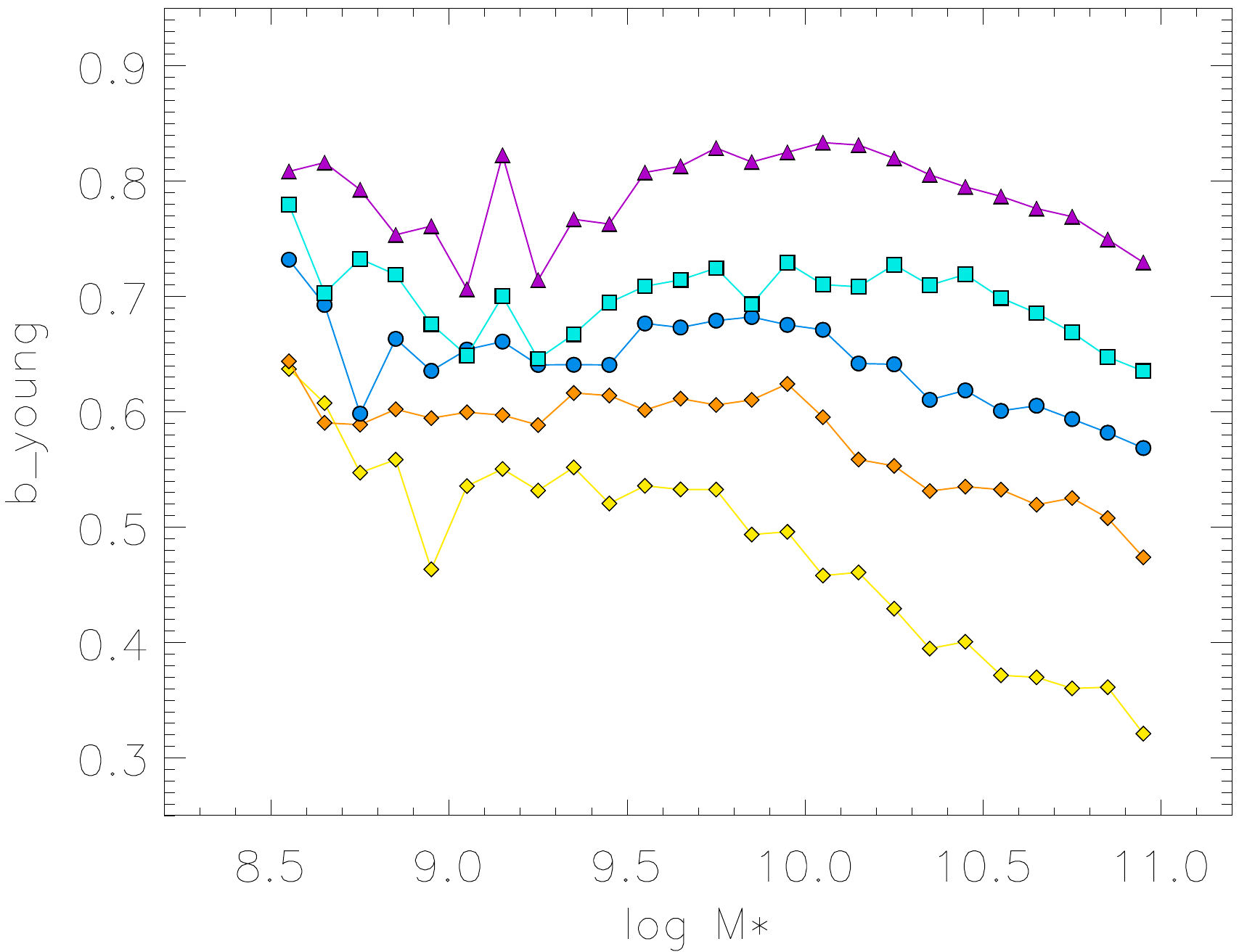}
    \includegraphics[scale=0.47]{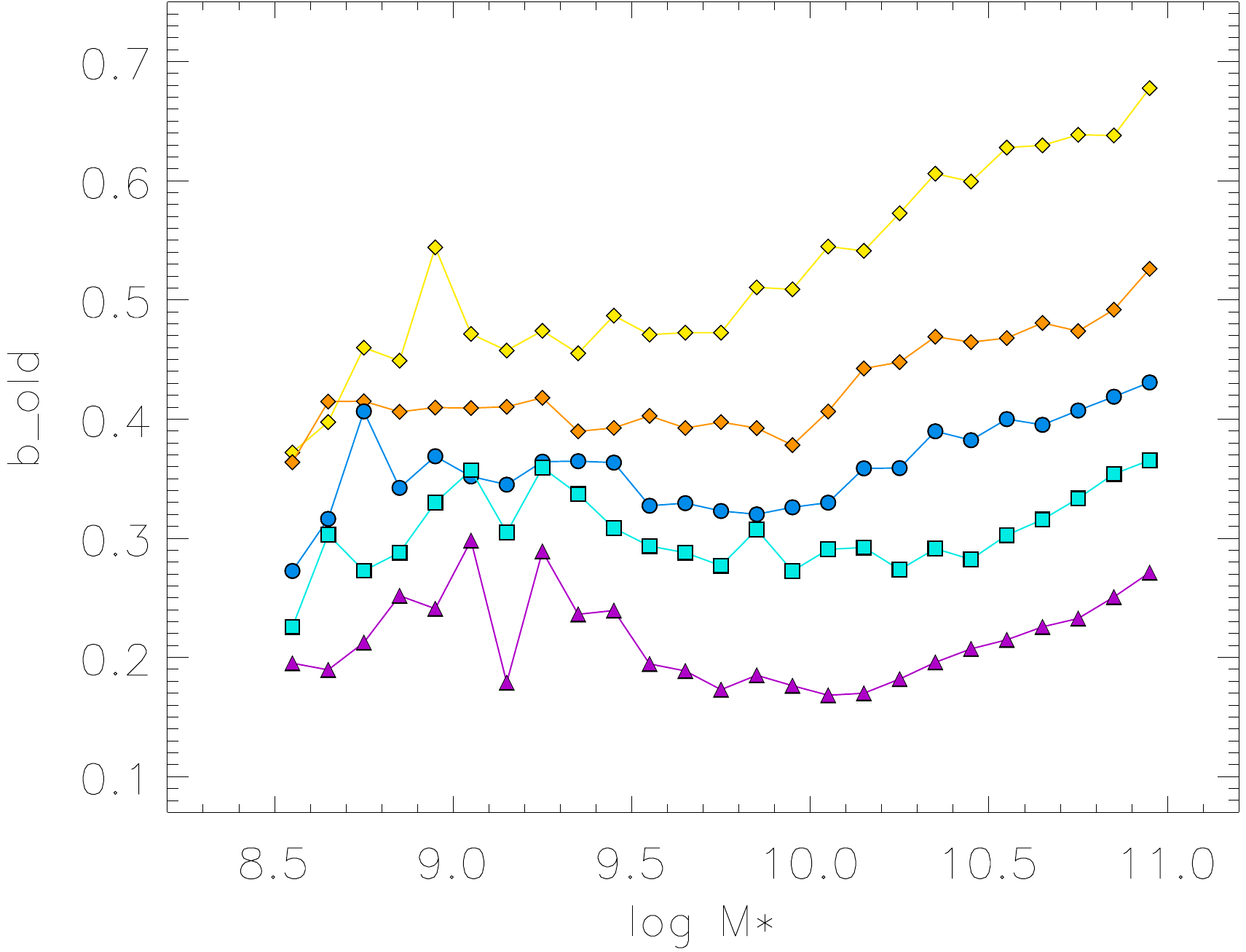}
  \end{center}
  \caption{
Top: average age t$_{av}$ (top figure) of the stellar population as a function of stellar mass in different bins of star formation rate. Middle and bottom: weight contributions of the young and old population b$_{young}$ and  b$_{old}$. The color code is the same as in Figure \ref{figure_18}. 
} \label{figure_19}
\end{figure}

\section{Initial Mass Function and Pre-Main Sequence Evolution}

For our calculation of SSB model spectra we have adopted a \cite{Chabrier2003} initial mass function (IMF). As discussed by \cite{Conroy2009}, there are viable alternatives such as the IMFs by \cite{Kroupa2001} and \cite{vanDokkum2008}. In order to investigate the influence of the choice of the IMF on our analysis we have additionally calculated model spectra with these two alternatives and then repeated the analysis. The resulting MZRs are shown in Figure \ref{figure_16} for high time resolution and t$_b$ = 0.1Myr. While the differences in metallicity arising from the choices of the different IMFs are noticeable, they are at most 0.05 dex and, thus, not a major source of systematic uncertainties. We have also included calculations with the original IMF suggested by \cite{Salpeter1955}, which is a simple power law without a modification at the low mass end. Here the differences are larger, 0.1 dex for the young and up to 0.2 dex for the old stellar population, respectively. Figure \ref{figure_16}  confirms the results found in previous work \citep{Cid_Fernandes2005, Wilkinson2017}.

The MESA stellar isochrones used in the calculation of our spectra include pre-main sequence evolution towards the zero-age mains sequence (ZAMS). As a consequence, young isochrones for which the more massive stars are leaving the main sequence have a contribution from stars with lower mass, which are still moving towards the main sequence and have luminosities significantly higher than the ZAMS at the same stellar mass (see \citealt{Choi2016} and \citealt{Dotter2016}). Because of the IMF these lower mass pre-main sequence objects could, thus, influence the integrated isochrone spectra and lead to differences when compared with isochrones just starting at the ZAMS. Since such isochrones have also been used in population synthesis diagnostic work, we have carried out an additional test with modified MESA isochrones, which start at the ZAMS. As it turns out, the MZRs obtained with model spectra calculated in this way are very similar to the ones where pre-main sequence evolution is included. The reason is the high luminosity of the more massive star on the main sequence or moving away, which have a dominating influence on the total spectrum. The inclusion of the pre-main sequence phase is obviously of minor importance for the results of our spectral diagnostics.

\begin{figure}[ht!]
  \begin{center}
    \includegraphics[scale=0.47]{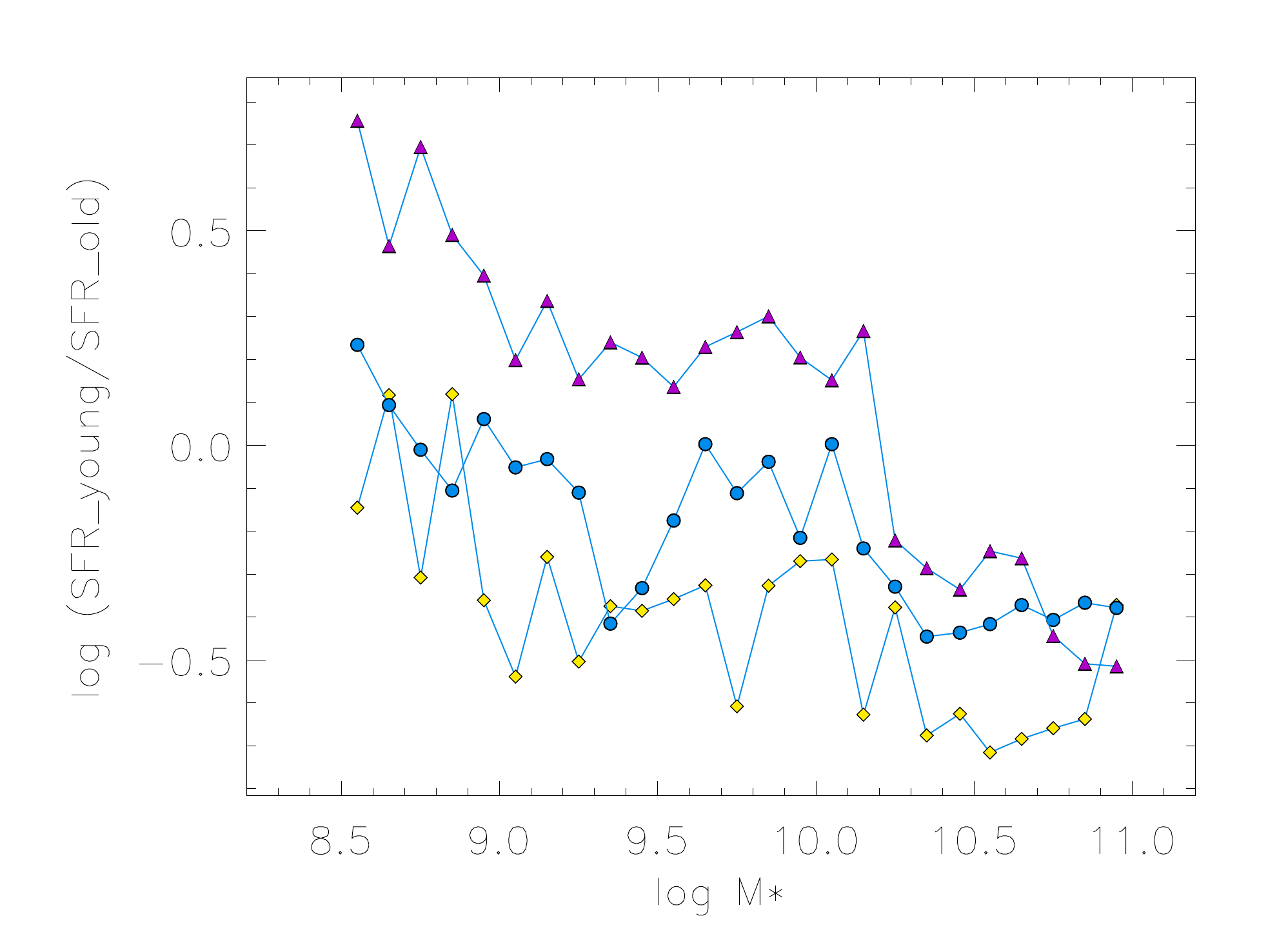}    
  \end{center}
  \caption{
The ratio of star formation rate of young to the old stellar population as a function of stellar mass for the galaxies in different bins of star formation rate. The color code is the same as in Figure \ref{figure_18} but only three bins are shown for the sake of clarity.
} \label{figure_20}
\end{figure}

\section{A Comparison with Lookback Galaxy Evolution Models}

Figure \ref{figure_7} and  \ref{figure_8} indicate a strong evolution of stellar metallicities as a function of age. The metallicities of the old population which is about 10 Gyr older are significantly lower and the difference is anti-correlated with galaxy stellar mass. This can be understood with the help of galaxy evolution models using the standard framework of chemical evolution. \cite{Kudritzki2021a,Kudritzki2021b} have introduced 'lookback' models where the competing processes of gas infall from the cosmic web and the circumgalactic medium, galactic winds and star formation are assumed to lead to a redshift dependent power law relationship between galactic gas mass and stellar mass, M$_g$ = A(z)M$_*^{\beta}$ and stars are formed along the observed star formation main sequences with star formation rates $\psi$ = $\psi_0$(z)M$^{\delta}$. With these assumptions and the usual chemical evolution equation metallicity [Z] becomes a simple analytical function of the ratio of stellar to gas mass M$_*$/M$_g$ and decreases when this ratio becomes smaller. This is very similar to the classical closed box model of chemical evolution \citep{Searle1972} but quantitatively the result is substantially different because of the power law dependence between gas and stellar mass.

Figure \ref{figure_17} (top) shows the metallicities of the lookback evolution model for each galaxy with different stellar mass for the young stars and stars 10 Gyr older overplotted to the result of our SDSS spectral analysis and the individual supergiants stars in galaxies. We note that these metallicities are luminosity weighted in the B-band. Details of the model are described in \cite{Kudritzki2021a,Kudritzki2021b}. We note that $\beta$ = 0.5 is adopted and $\psi_0$(z) has been increased by a factor of three to be in better agreement with the star formation rates of the SDSS galaxies at redshift z=0.08. The relative redshift dependence $\psi$(z) remains the same.

The agreement of the lookback models with the observations is good. The model reproduces the shape of the young stellar population MZR well and also the difference of the metallicities between the young stars and the old stars. We note that the galaxy stellar masses at lookback time $\Delta$t = 10 Gyr are significantly smaller than the final stellar masses. This is shown in the bottom part of Figure   \ref{figure_17}. The difference in stellar mass together with dependence of metallicity on M$_*$/M$_g$ is the reason why the decline of the metallicity of the old population with decreasing stellar mass is steeper than for the younger population. Because of this mass difference the red curve in Figure  \ref{figure_17} (top) does not represent a mass metallicity relationship at high redshift (z $\approx$ 2.5), but instead, describes the chemical evolution of these galaxies.

\section{Star Formation Rate Dependence of Metallicity and Ages}

From the study of HII region emission lines it has been found that MZRs contain a dependence on star formation rates as a third parameter \citep{Mannucci2010, Yates2012, Sanders2020}. Galaxies with higher star formation rates tend to have lower metallicities. Our analysis of stellar spectra provides the opportunity to investigate this effect with respect to stellar populations. For that purpose we follow the approach by \cite{Zahid2017} and divide the sample of our 200,000 SDSS spectra in quintiles of star formation rate. We then stack the spectra of these quintiles at each stellar mass and run our analysis on the stacked spectra of the quintiles. Since the spectra in the highest SFR quintile show strong HII emission lines for the lowest galaxy stellar masses, we include the effects of nebular continuum emission in our analysis as described in section 2. We use SSB with high age resolution and a short burst length t$_b$ = 0.1Myr for the analysis.

The resulting MZRs for the young and old population are given in Figure \ref{figure_18}. We see a clear correlation with SFR for both the young population at high masses and the old population at intermediate masses. The metallicity errors are of the same order as shown in Figure 7 and described in section 4.

We have also determined population ages t$_{young}$, t$_{old}$, and $t_{av}$. While t$_{young}$ and t$_{old}$ seem independent of star formation rate, the average age of the total population t$_{av}$ as shown in Figure \ref{figure_19}  reveals a very clear dependence. The average population in galaxies with a lower star formation rate is significantly older. Obviously, at the lower star formation rates it takes the galaxies much longer to build up their stellar population. In consequence, the contribution by the old stellar population b$_{old}$ to the spectral fit is much higher, as is also shown in Figure \ref{figure_19}. (Note that t$_{av}$ = b$_{young}$t$_{young}$ + b$_{old}$t$_{old}$ and that the second term dominates).

The stronger contribution of the older population is also reflected in the star formation history expressed by the ratio $\psi_{young}$/ $\psi_{old}$ of star formation rates of the young to the old population. As Figure \ref{figure_20} demonstrates, this ratio is significantly smaller for galaxies with presently lower star formation rates.


 \section{Summary and Conclusions}

In this section we summarize and discuss the major results of this work. In a first step we have investigated the influence of star formation burst lengths t$_b$ on the results of the population synthesis analysis. As observations show (see section 3), t$_b$ depends on the size of the star forming region and can range from a fraction of a Myr to many Myr. We find that the assumption of burst length influences the spectra of stellar isochrones significantly depending, of course, on isochrone age t and t$_b$. For ages t with t$_b$/t $\le$ 0.02 the signatures of burst length are clearly visible in the spectra. This affects the results of the population synthesis analysis, in particular, the metallicities of the young stellar population. Judging the quality of the spectral fits by means of their minimum $\chi^2$ value we find that burst spectra calculated with t$_b$ = 0.1 and 1.0 Myr give the best fit. Fits with 10 Myr bursts are worse.

Since we model the spectrum of the integrated stellar population as a sequence of bursts with different ages, the time resolution of this sequence is important. We have, therefore, tested the influence of the time resolution using two grids with different resolution of isochrone age steps. We find that the higher time resolution leads to lower $\chi^2$ values of the spectral fits.

The analysis of the galaxy spectra with respect to metallicity and age requires a determination of interstellar reddening and extinction. Since it is well know that in the star forming regions of galaxies reddening laws deviate from the standard law of the diffuse ISM characterized by the  R$_V$ = 3.1, we determine R$_V$ as an independent additional parameter. We find values clearly larger than 3.1 indicating a strong contribution of dense star forming regions to the reddening. The  R$_V$ values are largest for low mass galaxies. For these galaxies the average age of the young population is lowest. In consequence, the contribution by stars originating from compact regions of star formation might be strongest. Allowing for deviations of R$_V$ from 3.1 has a significant influence on the determination of metallicity (up to 0.4 dex), population ages (10 to 50\% and, of course, interstellar extinction (0.2 magnitudes). We note that this striking general difference from the standard reddening law in star forming galaxies has important consequences for other aspects of astrophysics such as, for instance,  distance determinations and the cosmological distance ladder or the estimate of galaxy luminosities.

The metallicity of the young population of our star forming SDSS galaxies is in good agreement with massive supergiant stars where the metallicities are obtained through accurate spectral analysis of a large sample of galaxies in the local universe. We take this as a crucial confirmation that our population synthesis analysis technique is reliable. The average age we derive for the young population is between 50 and 300 Myrs increasing with galactic stellar mass. It is, thus, in the same range as the supergiant ages and, consequently, we expect the metallicities to be similar.

We also determine the average metallicities and ages of the old population. The metallicities are significantly smaller than for the young population. The age difference to the young population is about 10 Gyr. Thus, we are seeing the results of chemical evolution over 10 Gyr. We compare these results with galaxy evolution look back models and find good agreement. Both, the mass-metallicity relationship of the young population and the metallicities of the old population are well reproduced. This supports a galaxy evolution model where metallicity depends on the ratio of gas mass to stellar mass and where this ratio decreases with time.

For this comparison we calculated B-band luminosity weighted metallicities of our models because our observed and model spectra are normalized to unity between 4400 to 4450 \angstrom\ and our analysis effectively provides B-band luminosity weighted metallicities and ages. However, we have checked the effects of normalization by introducing normalizations between 5500 and 5550  and 6950 and 7000 \angstrom\ and found that they are small. Changes due to the different normalizations are comparable to the fit uncertainties.

From the weight contributions to the total spectrum of isochrones at different ages we can also estimate star formation rates. We calculate the ratio of star formation rate of the young and old population and find an anti-correlation with galaxy stellar mass. Low mass galaxies show a much higher star formation contribution of the younger population. On the other hand, more massive galaxies are dominated by stars that formed at early times. This is in agreement with the observations of ``galaxy downsizing'' as originally detected by \cite{Cowie1996} and followed up in comprehensive galaxy survey work (see, for instance, \citealt{Piliyugin2011, Thomas2019} and references therein).

For the calculation of our model spectra we have assumed a \cite{Chabrier2003} initial mass function. In order to test the influence of this assumption we repeat the analysis using the IMFs by \cite{Kroupa2001}, \cite{vanDokkum2008} and \cite{Salpeter1955}. The differences with respect to metallicity are very small ($\leq$ 0.05 dex) for the first three IMFs, but are larger for the Salpeter IMF (0.1 dex and 0.2 dex for the young and old population, respectively).

Finally, we investigate the dependence of metallicity and ages on star formation rate. We stack the spectra of the 200,000 SDSS galaxies in quintiles of SFR at each stellar mass and repeat the analysis for each quintile. We find a correlation of metallicity with SFR for higher and intermediate stellar masses for the young and old population. In addition, the average age of the total stellar population shows a clear dependence on SFR. The population in galaxies with lower SFR is significantly older and the contribution of the old population to the integrated spectrum is much larger. In the same way, the ratio of SFR of the young to the old population is smaller for galaxies with smaller SFR.

In summary, the results provide important implications for galaxy evolution and underline the power of spectroscopic population synthesis analysis techniques.

\acknowledgments
This work was initiated  and supported by the Munich Excellence Cluster Origins funded by the Deutsche Forschungsgemeinschaft (DFG, German Research Foundation) under Germany's Excellence Strategy EXC-2094 390783311. It is a pleasure to thank our colleague Andi Burkert for inspiring and stimulating discussion along this project.\\

\bibliography{ms}

\end{document}